\documentclass[aps,twocolumn,prd,superscriptaddress]{revtex4-2}

\usepackage[utf8]{inputenc}

\usepackage{mathtools}
\usepackage{amsfonts}
\usepackage{mathrsfs}
\usepackage{bbm}
\usepackage{slashed}
\usepackage{nicefrac}

\usepackage{graphicx}
\usepackage{color}
\usepackage{array}

\usepackage{placeins}
\usepackage{booktabs}
\usepackage{makecell}
\usepackage{floatrow}

\usepackage{xspace}
\usepackage{siunitx}
\usepackage{hyperref}
\usepackage[nameinlink]{cleveref}
\usepackage{appendix}

\usepackage{xifthen}
\usepackage{xcolor}
\hypersetup{
	colorlinks,
	linkcolor={red!75!black},
	citecolor={blue!75!black},
	urlcolor={blue!75!black}
}

\newcommand{\tinytext}[1]{\text{\tiny{#1}}}

 \newcommand{\gettitle}{Spectral Reconstruction with Deep Neural Networks}
 
 \hypersetup{
 	pdftitle={\gettitle},
 	pdfauthor={Kades,Pawlowski,Rothkopf,Scherzer,Urban,Wetzel,Wink,Ziegler},
 	pdfkeywords={spectral function}
 	{correlations functions} {real time}
 	{machine learning} {spectral reconstruction},
 	bookmarksopen=true,
 	bookmarksopenlevel=2,
 	bookmarksnumbered=true
 }
  
\begin{document}

\title{Spectral Reconstruction with Deep Neural Networks}

\newcommand{\getHeidelbergAffiliation}{\affiliation{Institut f\"ur Theoretische Physik, Universit\"at Heidelberg, Philosophenweg 16, D-69120 Heidelberg, Germany}}
\newcommand{\getStavangerAffiliation}{\affiliation{Faculty of Science and Technology, University of Stavanger, NO-4036 Stavanger, Norway}}
\newcommand{\getEMMIAffiliation}{\affiliation{ExtreMe Matter Institute EMMI, GSI, Planckstr.~1, D-64291 Darmstadt, Germany}}
\newcommand{\getPerimeterAffiliation}{\affiliation{Perimeter Institute for Theoretical Physics, Waterloo, Ontario, Canada N2L 2Y5}}

\author{Lukas~Kades} \getHeidelbergAffiliation
\author{Jan~M.~Pawlowski} \getHeidelbergAffiliation \getEMMIAffiliation
\author{Alexander~Rothkopf} \getStavangerAffiliation
\author{Manuel~Scherzer} \getHeidelbergAffiliation
\author{Julian~M.~Urban} \getHeidelbergAffiliation
\author{Sebastian~J.~Wetzel} \getHeidelbergAffiliation \getPerimeterAffiliation
\author{Nicolas~Wink} \getHeidelbergAffiliation
\author{Felix~P.G.~Ziegler} \getHeidelbergAffiliation
 
%%%%%%%%%%%%%%%%%%%%%%%%%%%%%%%%

\begin{abstract}
We explore artificial neural networks as a tool for the reconstruction
of spectral functions from imaginary time Green's functions, a classic
ill-conditioned inverse problem. Our ansatz is based on a supervised
learning framework in which prior knowledge is encoded in the training
data and the inverse transformation manifold is explicitly
parametrised through a neural network. We systematically investigate
this novel reconstruction approach, providing a detailed analysis of
its performance on physically motivated mock data, and compare it to
established methods of Bayesian inference. The reconstruction accuracy
is found to be at least comparable, and potentially superior in
particular at larger noise levels. We argue that the use of labelled
training data in a supervised setting and the freedom in defining an
optimisation objective are inherent advantages of the present approach
and may lead to significant improvements over state-of-the-art methods
in the future. Potential directions for further research are discussed
in detail.
\end{abstract}

\maketitle

%%%%%%%%%%%%%%%%%%%%%%%%%%%%%%%%

\section{Introduction}\label{sec:intro}

Machine Learning has been applied to a variety of problems in the
natural sciences. For example, it is regularly deployed in the
interpretation of data from high-energy physics detectors
\cite{Guest:2018yhq, Radovic:2018dip}. Algorithms based on learning
have shown to be highly versatile, with their use extending far beyond
the original design purpose. In particular, deep neural networks have
demonstrated unprecedented levels of prediction and
generalisation performance, for reviews see e.g. \cite{LeCun2015,
	1404.7828}. Machine Learning architectures are also
increasingly deployed for a variety of problems in the theoretical
physics community, ranging from the identification of phases and order
parameters to the acceleration of lattice simulations \cite{nphys4035,
	PhysRevD.97.094506, science.aag2302, PhysRevE.96.022140,
	PhysRevB.96.184410, PhysRevB.94.195105, PhysRevE.95.062122,
	PhysRevB.95.035105, PhysRevB.95.041101, Karpie:2019eiq, 1811.03533, Bluecher:2020kxq}.

Ill-conditioned inverse problems lie at the heart of some of the most
challenging tasks in modern theoretical physics. One pertinent example
is the computation of real-time properties of strongly correlated
quantum systems. Take e.g. the phenomenon of energy and charge
transport, which so far has defied a quantitative understanding from
first principles. This universal phenomenon is relevant to systems at
vastly different energy scales, ranging from ultracold quantum gases
created with optical traps to the quark-gluon plasma born out of
relativistic heavy-ion collisions.

While static properties of strongly correlated quantum systems are by
now well understood and routinely computed from first principles, a
similar understanding of real-time properties is still subject to
ongoing research. The thermodynamics of strongly coupled systems, such
as the quark gluon plasma, has been explored using the combined
strength of different non-perturbative approaches, such as functional
renormalisation group methods and lattice field theory calculations.
There are two limitations affecting most of these approaches: Firstly,
in order to carry out quantitative computations, time has to be
analytically continued into the complex plane, to so-called Euclidean
time. Secondly, explicit computations are either fully numerical or at
least involve intermediate numerical steps.

This leaves us with the need to eventually undo the analytic
continuation of Euclidean correlation functions, which are known only
approximately. The most relevant examples are two-point functions, the
so-called Euclidean propagators. The spectral representation of
quantum field theory relates the propagators, be they in Minkowski or
Euclidean time, to a single function encoding their physics, the
so-called spectral function. The number of different structures
contributing to a spectral function are in general quite limited and
consist of poles and cuts. If we can extract from the Euclidean
two-point correlator its spectral function, we may immediately compute
the corresponding real-time propagator.

If we know the Euclidean propagator analytically, this information
allows us in principle to recover the corresponding Minkowski time
information. In practice, however, the limitation of having to
approximate correlator data (e.g. through simulations) turns the
computation of spectral functions into an ill-conditioned problem. The
most common approach to give meaning to such inverse problems is
Bayesian inference. It incorporates additional prior domain knowledge
we possess on the shape of the spectral function to regularise the
inversion task. The positivity of hadronic spectral functions is one
prominent example. The Bayesian approach has seen continuous
improvement over the past two decades in the context of spectral
function reconstructions. While originally it was restricted to
maximum a posteriori estimates for the most probable spectral function
given Euclidean data and prior information
\cite{Jarrell:1996rrw,Asakawa:2000tr,Burnier:2013nla}, in its most
modern form it amounts to exploring the full posterior distribution
\cite{Rothkopf:2019dzu}. An important aspect of the work is to develop
appropriate mock data tests to benchmark the reconstruction
performance before applying it to actual data. Generally, the success
of a reconstruction method stands or falls with its performance on
physical data. While this seems evident, it was in fact a hard lesson
learned in the history of Bayesian reconstruction methods, a lesson
which we want to heed.

Inverse problems of this type have also drawn quite some attention in
the machine learning (ML) community \cite{1802.08406, 1803.00092,
	1805.07281, Rother2018}. In the present work we build upon both the
recent progress in the field of ML, particularly deep learning, as
well as results and structural information gathered in the past
decades from Bayesian reconstruction methods. We set out to isolate a
property of neural networks that holds the potential to improve upon
the standard Bayesian methods, while retaining their advantages,
utilising the already gathered insight in their study.

\begin{figure*}
	\includegraphics[width=1.0\textwidth]{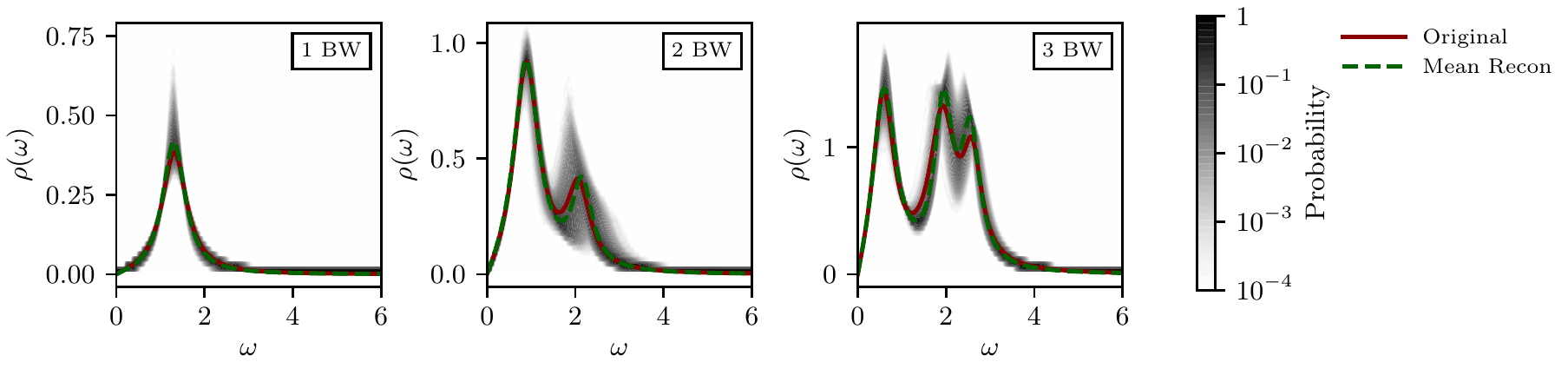}
	\caption{Examples of mock spectral functions reconstructed via our
	neural network approach for the cases of one, two and three
	Breit-Wigner peaks. The chosen functions mirror the desired locality
	of suggested reconstructions around the original function (red line).
	Additive, Gaussian noise of width $10^{-3}$ is added to the
	discretised analytic form of the associated propagator of the same
	original spectral function multiple times. The shaded area depicts
	for each frequency $\omega$ the distribution of resulting outcomes,
	while the dashed green line corresponds to the mean. The results are
	obtained from the FC parameter network optimised with the parameter
	loss. The network is trained on the largest defined parameter space
	which corresponds to the volume Vol O. The uncertainty for
	reconstructions decreases for smaller volumes as illustrated in
	\Cref{fig:volumeimpact}. A detailed discussion on the properties and
	problems of a neural network based reconstruction is given in
	\Cref{sec:reconneuralnetwork}.}
	\label{fig:numbwimpact}
\end{figure*}

Consider a feed-forward deep neural network that takes Euclidean
propagator data as input and outputs a prediction of the associated
spectral function. Although the reasoning behind this ansatz is rather
different, one can draw parallels to more traditional methods. In the
Bayesian approach, prior information is explicitly encoded in a prior
probability functional and the optimisation objective is the precise
recovery of the given propagator data from the predicted spectral
function. In contrast, the neural network based reconstruction is
conditioned through supervised learning with appropriate training
data. This corresponds to implicitly imposing a prior distribution on
the set of possible predictions, which, as in the Bayesian case,
regularises the reconstruction problem. Optimisation objectives are
now expressed in terms of loss functions, allowing for greater
flexibility. In fact, we can explicitly provide pairs of correlator
and spectral function data during the training. Hence, not only can we
aim for the recovery of the input data from the predictions as in the
Bayesian approach, but we are now also able to formulate a loss
directly on the spectral functions themselves. This constitutes a much
stronger restriction on potential solutions for individual
propagators, which could provide a significant advantage over other
methods. The possibility to access all information of a given sample
with respect to its different representations also allows the
exploration of a much broader set of loss functions, which could
benefit not only the neural network based reconstruction, but also
lead to a better understanding and circumvention of obstacles related
to the inverse problem itself. Such an obstacle is given, for example,
by the varying severity of the problem within the space of considered
spectral functions. By employing adaptive losses, inhomogeneities of
this type could be neutralised.

Similar approaches concerning spectral functions that consist of
normalised sums of Gaussian peaks have already been discussed in
\cite{Fournier2018, Yoon2018}. In this work, we investigate the
performance of such an approach using mock data of physical resonances
motivated by quantum field theory, and compare it to state-of-the-art
Bayesian methods. The data are given in the form of linear
combinations of unnormalised Breit-Wigner peaks, whose distinctive
tail structures introduce additional difficulties (see
\Cref{fig:numbwimpact} for an example reconstruction). Using only a
rather naive implementation, the performance of our ansatz is
demonstrated to be at least comparable and potentially superior,
particularly for large noise levels. We then discuss potential
improvements of the architecture, which in the future could establish
neural networks to a state-of-the-art approach for accurate
reconstructions with a reliable estimation of errors.

The paper is organised as follows. The spectral reconstruction problem
is defined in \Cref{sec:specreca}. State-of-the-art Bayesian
reconstruction methods are summarised in \Cref{sec:specrecb}. In
\Cref{sec:specrecc} we discuss the application of neural networks and
potential advantages. \Cref{sec:recprescr} contains details on the
design of the networks and defines the optimisation procedure.
Numerical results are presented and compared to Bayesian methods in
\Cref{sec:results}. We summarise our findings and discuss future work
in \Cref{sec:conclusion}.

%%%%%%%%%%%%%%%%%%%%%%%%%%%%%%%%%%%%%%%%%%%%%%%%%%%%%%%%%%%%%%%%%%%%%%%%
\section{Spectral reconstruction and potential advantages} \label{sec:theory}
%%%%%%%%%%%%%%%%%%%%%%%%%%%%%%%%%%%%%%%%%%%%%%%%%%%%%%%%%%%%%%%%%%%%%%%%

%%%%%%%%%%%%%%%%%%%%%%%%%%%%%%%%%%%%%%%%%%%%%%%%%%%%%%%%%%%%%%%%%%%%%%%%
\subsection{Defining the problem} \label{sec:specreca}
%%%%%%%%%%%%%%%%%%%%%%%%%%%%%%%%%%%%%%%%%%%%%%%%%%%%%%%%%%%%%%%%%%%%%%%%

Typically, correlation functions in equilibrium quantum field theories
are computed in imaginary time after a Wick rotation $t\to i t \equiv
\tau$, which facilitates both analytical and numerical computations.
In strongly correlated systems, a numerical treatment is in most cases
inevitable. Such a setup leaves us with the task to reconstruct
relevant information, such as the spectrum of the theory, or genuine
real-time quantities such as transport coefficients, from the
Euclidean data.

The information we want to access is encoded in the associated
spectral function $\rho$. For this purpose it is most convenient to
work in momentum space both for $\rho$ and the corresponding
propagator $G$. The relation between the Euclidean propagator and the
spectral function is given by the well known K{\"a}llen-Lehmann
spectral representation,
\begin{align} \label{eq:KLrep}
G(p)&=\int_{0}^{\infty} \frac{\mathrm{d}\omega}{\pi}
\frac{\omega\,\rho(\omega)}{\omega^2+p^2} \equiv \int_{0}^{\infty}
\mathrm{d}\omega\, K(p,\omega) \rho(\omega) \, ,
\end{align}
which defines the corresponding K{\"a}llen-Lehmann kernel. The
propagator is usually only available in the form of numerical data,
with finite statistical and systematic uncertainties, on a discrete
set of $N_p$ points, which we abbreviate as $G_i = G(p_i)$. The most
commonly used approach is to work directly with a discretised version
of \labelcref{eq:KLrep}. We utilise the same abbreviation for the
spectral function, i.e.\ $\rho_i = \rho(\omega_i)$, discretised on
$N_\omega$ points. This lets us state the discrete form of
\labelcref{eq:KLrep} as
\begin{align}
\label{eq:grid_discretized_version} G_i=\sum_{j=1}^{N_\omega} K_{ij}\,
\rho_j\, \, ,
\end{align}
where $K_{ij} = K(p_i,\omega_j) \Delta\omega_j $. This amounts to a
classic ill-conditioned inverse problem, similar in nature to those
encountered in many other fields, such as medical imaging or the
calibration of option pricing methods. Typical errors on the input
data $G(p_i)$ are on the order of $10^{-2}$ to $10^{-5}$ when the
propagator at zero momentum is of the order of unity.

To appreciate the problems arising in such a reconstruction more
clearly, let us assume we have a suggestion for the spectral function
$\rho_{\text{\tiny{sug}}}$ and its corresponding propagator
$G_\tinytext{sug}$. The difference to the measured data is encoded in
\begin{align} \label{eq:normG}
\| G(p) - &G_{\tinytext{sug}}(p)\| =
\nonumber \\[1ex] & \bigg \| \int_0^\infty \frac{\mathrm{d}\omega}{\pi}
\frac{\omega}{\omega^2+p^2} \, \Big[\rho(\omega) -
\rho_{\tinytext{sug}}(\omega)\Big] \bigg\| \, ,
\end{align}
with a suitable norm $\|.\|$. Evidently, even if this expression
vanishes point-wise, i.e.\
$\| G(p_i) - G_{\tinytext{sug}}(p_i)\| = 0 $ for all $p_i$, the
spectral function is not uniquely fixed. Experience has shown that
with typical numerical errors on the input data, qualitatively very
different spectral functions can lead to in this sense equivalent
propagators. This situation can often be improved on by taking more
prior knowledge into account, c.f.\ the discussion in
\cite{Cyrol:2018xeq}. This includes properties such as:

\begin{enumerate}
	\item Normalisation and positivity of spectral functions of
	asymptotic states. For gauge theories, this may reduce to just the
	normalisation to zero, expressed in terms of the Oehme-Zimmermann
	superconvergence \cite{Oehme:1979bj, Oehme:1990kd}.
	\item Asymptotic behaviour of the spectral function at low and high
	frequencies.
	\item The absence of clearly unphysical features, such as
	drastic oscillations in the spectral function and the propagator.
\end{enumerate}

Additionally, the parametrisation of the spectral function in terms of
frequency bins is just one particular basis. In order to make
reconstructions more feasible, other choices, and in particular
physically motivated ones, may be beneficial, c.f.\ again the
discussion in \cite{Cyrol:2018xeq}. In this work, we consider a basis
formulated in terms of physical resonances, i.e.\ Breit-Wigner peaks.

%%%%%%%%%%%%%%%%%%%%%%%%%%%%%%%%%%%%%%%%%%%%%%%%%%%%%%%%%%%%%%%%%%%%%%%%
\subsection{Existing methods} \label{sec:specrecb}
%%%%%%%%%%%%%%%%%%%%%%%%%%%%%%%%%%%%%%%%%%%%%%%%%%%%%%%%%%%%%%%%%%%%%%%%

The inverse problem as defined in \labelcref{eq:KLrep} has an exact
solution in the case of exactly known, discrete correlator data
\cite{Cuniberti:2001hm}. However, as soon as noisy inputs are
considered, this approach turns out to be impractical
\cite{Burnier:2011jq}. Therefore, the most common strategy to treat
this problem is via Bayesian inference. This approach is based on
Bayes' theorem, which states that the posterior probability is
essentially given by two terms, the likelihood function and prior
probability:
\begin{align}
P(\rho|D,I) \propto P(D|\rho,I) \, P(\rho|I) \, .
\end{align}
It explicitly includes additionally available prior information on the
spectral function in order to regularise the inversion task. The
likelihood $P(D|\rho)$ encodes the probability for the input data $D$
to have arisen from the test spectral function $\rho$, while $P(\rho)$
quantifies how well this test function agrees with the available prior
information. The two probabilities fully specify the posterior
distribution in principle, however they may be known only implicitly.
In order to gain access to the full distribution, one may sample from
the posterior, e.g.\ through a Markov Chain Monte Carlo process in the
parameter space of the spectral function. However, in practice one is
often content with the maximum a posteriori (MAP) solution. Given a
uniform prior, the Maximum Likelihood Estimate (MLE) corresponds to an
estimate of the MAP.

%%%%%%%%%%%%%%%%%%%%%%%%%%%%%%%%%%%%%%%%%%%%%%%%%%%%%%%%%%%%%%%%%%%%%%%%
\subsection{Advantages of neural networks} \label{sec:specrecc}
%%%%%%%%%%%%%%%%%%%%%%%%%%%%%%%%%%%%%%%%%%%%%%%%%%%%%%%%%%%%%%%%%%%%%%%%

In order to make genuine progress, we set out in this study to explore
methods in which our prior knowledge of the analytic structure can be
encoded in different ways. To this end, our focus lies on the use of
Machine Learning in the form of artificial neural networks. These
feature a high flexibility in the encoding of information by learning
abstract internal representation. They possess the advantageous
properties that prior information can be explicitly provided through
the training data, and that the solution space can be regularised by
choosing appropriate loss functions.

Minimising \labelcref{eq:normG}, while respecting the constraints
discussed in \Cref{sec:specreca}, can be formulated as minimising the
propagator loss
\begin{align}
\label{eq:LGrho}
	L_G(\rho_\tinytext{sug}) &=\lVert
	G[\rho_\tinytext{sug}]-G[\rho]\rVert
\, .
\end{align}
This corresponds to indirectly working on a norm or loss function for
$\rho$, the spectral function loss
\begin{align}
\label{eq:Lrho}
	L_\rho(\rho_\tinytext{sug}) = \lVert
	\rho_\tinytext{sug} - \rho\rVert
\, .
\end{align}
Of course, the optimisation problem as given by \labelcref{eq:Lrho} is
intractable, since it requires the knowledge of the true spectral
function $\rho$. Minimising $L_\rho(\rho_\tinytext{sug})$ for a given
set of $\{\rho_{\text{\tiny{sug}}}\}$ also minimises $L_G$, since the
K{\"a}ll{\'e}n–Lehmann representation \labelcref{eq:KLrep} is a linear
map. In turn, however, minimising $L_G$ does not uniquely determine
the spectral function, as has already been mentioned. Accordingly, the
key to optimise the spectral reconstruction is the ideal use of all
known constraints on $\rho$, in order to better condition the problem.
The latter aspect has fueled many developments in the area of spectral
reconstructions in the past decades.

Given the complexity of the problem, as well as the interrelation of
the constraints, this calls, in our opinion, for an application of
supervised machine learning algorithms for an optimal utilisation of
constraints. To demonstrate our reasoning, we generate a training set
of known pairs of spectral functions and propagators and train a
neural network to reconstruct $\rho$ from $G$ by minimising a suitable
norm, utilising both $L_G$ and $L_\rho$ during the training. When the
network has converged, it can be applied to measured propagator data
$G$ for which the corresponding $\rho$ is unknown.

Estimators learning from labelled data provide a potentially
significant advantage due to the employed supervision, because the
loss function is minimised a priori for a whole range of possible
input/output pairs. Accordingly, a neural network aims to learn the
entire set of inverse transformations for a given set of spectral
functions. After this mapping has been approximated to a sufficient
degree, the network can be used to make predictions. This is in
contrast to standard Bayesian methods, where the posterior
distribution is explored on a case by case basis. Both approaches may
also be combined, e.g.\ by employing a neural network to suggest a
solution $\rho_{sug}$, which is then further optimised using a
traditional algorithm.

The given setup forces the network to regularise the ill-conditioned
problem by reproducing the correct training spectrum in accord with
our criteria for a successful reconstruction. It is the inclusion of
the training data and the free choice of loss functions that allows
the network to fully absorb all relevant prior information. This
ability is an outstanding property of supervised learning methods,
which could yield potentially significant improvements over existing
frameworks. for such constraints are the analytic structure
of the propagator, asymptotic behaviors and normalisation constraints.

The parametrisation of an infinite set or manifold of inverse
transformations by the neural network also enables the discovery of
new loss functions which may be more appropriate for a reliable
reconstruction. This includes, for example, the exploration of
correlation matrices with adapted elements, in order to define a
suitable norm for the given and suggested propagators. Existing,
iterative methods may also benefit from the application of such
adaptive loss functions. These may include parameters, point-like
representations and arbitrary other characteristics of a given
training sample.

Formulated in a Bayesian language, we set out to explicitly train the
neural network to predict MAP estimates for each possible input
propagator, given the training data as prior information. By salting
the input data with noise, the network learns to return a denoised
representation of the associated spectral functions.

\section{A neural network based reconstruction} \label{sec:recprescr}

Neural networks provide high flexibility with regard to their
structure and the information they can process. They are capable of
learning complex internal representations which allow them to extract
the relevant features from a given input. A variety of network
architectures and loss functions can be implemented in a
straightforward manner using modern Machine Learning frameworks. Prior
information can be explicitly provided through a systematic selection
of the training data. The data itself provides, in addition to the
loss function, a regularisation of possible suggestions. Accordingly,
the proposed solutions have the advantage to be similar to the ones in
the training data.

The section starts with notes on the design of the neural networks we
employ and ends with a detailed introduction of the training procedure
and the utilised loss functions.

\subsection{Design of the neural networks}

\begin{figure}
	\includegraphics[width=\columnwidth]{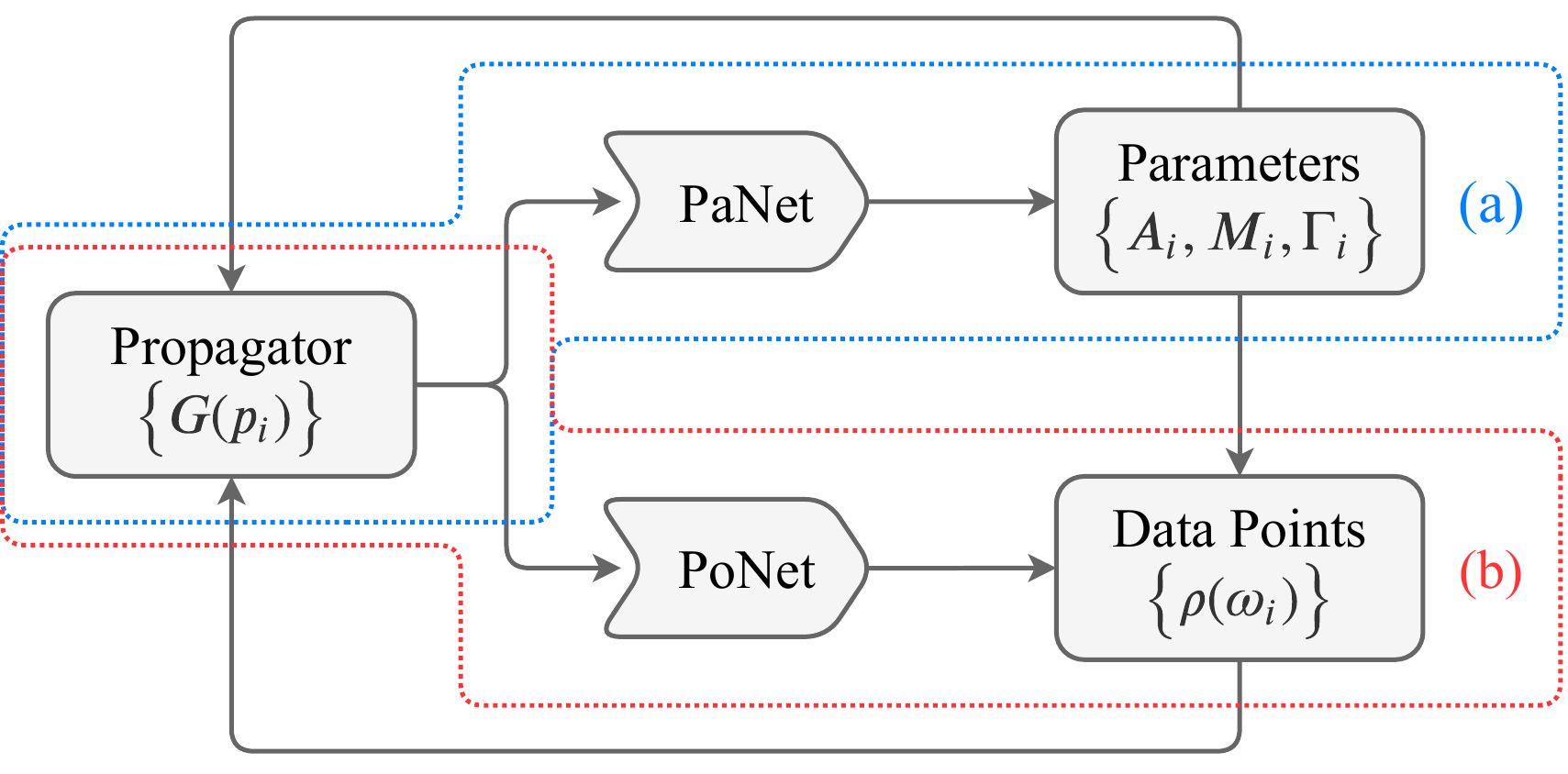}
	\caption{Sketch of our strategy for reconstructing (a) the parameters
	using the PaNet and (b) the discretised data points using the PoNet
	(and by extension also the PoNetVar). Details on the architectures
	are given in \Cref{sec:mocktst}.}
	\label{schematic}
\end{figure}

We construct two different types of deep feed-forward neural networks.
The input layer is fed with the noisy propagator data $G(p)$. The
output for the first type is an estimate of the parameters of the
associated $\rho$ in the chosen basis, which we denote as
\textit{parameter net (PaNet)}. For the second type, the network is
trained directly on the discretised representation of the spectral
function. This network will be referred to as \textit{point net
  (PoNet)}. A consideration of a variable number of Breit-Wigners is
feasible per construction by the point-like representation of the
spectral function within the output layer. This kind of network will
in the following be abbreviated by \textit{PoNetVar}. See
\Cref{schematic} for a schematic illustration of our strategy. Note
that in all cases a basis for the spectral function is provided either
explicitly through the structure of the network or implicitly through
the choice of the training data. If not stated otherwise, the
numerical results presented in the following always correspond to
results from the PaNet.

We compare the performance of fully-connected (FC) and convolutional
(Conv) layers as well as the impact of their depth and width. In
general, choosing the numbers of layers and neurons is a trade-off
between the expressive power of the network, the available memory and
the issue of overfitting. The latter strongly depends on the number of
available training samples w.r.t.\ the expressivity. For fully
parametrised spectral functions, new samples can be generated very
efficiently for each training epoch, which implies an, in principle,
infinite supply of data. Therefore, in this case, the risk of
overfitting is practically non-existent. The specific dimensions and
hyperparameters used for this work are provided in
\Cref{sec:mocktst}. Numerical results can be found in
\Cref{sec:results}.

\subsection{Training strategy}
\label{sec:trainingstrategy}

The neural network is trained with appropriately labelled input data
in a supervised manner. This approach allows to implicitly impose a
prior distribution in the Bayesian sense. The challenge lies in
constructing a training dataset that is exhaustive enough to contain
the relevant structures that may constitute the actual spectral
functions in practical applications.

From our past experience with hadronic spectral functions in lattice
QCD and the functional renormalisation group, the most relevant
structures are peaks of the Breit-Wigner type, as well as thresholds.
The former present a challenge from the point of view of inverse
problems, as they contain significant tail contributions, contrary
e.g.\ to Gaussian peaks, which approach zero exponentially fast.
Thresholds on the other hand set in at finite frequencies, often
involving a non-analytic kink behavior. In this work, we only consider
Breit-Wigner type structures as a first step for the application of
neural networks to this family of problems.

Mock spectral functions are constructed using a superposed collection of
Breit-Wigner peaks based on a parametrisation obtained directly from
one-loop perturbative quantum field theory. Each individual
Breit-Wigner is given by
\begin{align}
\rho^{(\tiny{BW})}(\omega) = \frac{4 A \Gamma
	\omega}{(M^2+\Gamma^2-\omega^2)^2 + 4 \Gamma^2 \omega^2} \,.
\label{eq:breitwigner}
\end{align}
Here, $M$ denotes the mass of the corresponding state, $\Gamma$ its
width and $A$ amounts to a positive normalisation constant.

Spectral functions for the training and test set are constructed from
a combination of at most $N_\textnormal{BW}=3$ different Breit-Wigner
peaks. Depending on which type of network is considered, the Euclidean
propagator is obtained either by inserting the discretised spectral
function into \labelcref{eq:grid_discretized_version}, or by a
computation of the propagator's analytic representation from the given
parameters. The propagators are salted both for the training and test
set with additive, Gaussian noise
\begin{align}
G_i^{\rm noisy}=G_i+\epsilon\,.
\end{align}
This is a generic choice which allows to quantify the performance of
our method at different noise levels.

The advantage of neural networks to have direct access to different
representations of a spectral function implies a free choice of
objective functions in the solution space. We consider three simple
loss functions and combinations thereof. The (pure) propagator loss
$L_G(\rho_\tinytext{sug})$ defined in \labelcref{eq:LGrho} represents
the most straightforward approach. This objective function is
accessible also in already existing frameworks, such as Bayesian
Reconstruction (BR) or Hamiltonian Monte Carlo (HMC) methods, in
particular the GrHMC framework (referring to the retarded propagator
$G_r$) developed in \cite{Cyrol:2018xeq}. It is implemented in this
work to facilitate a quantitative comparison. In contrast, the loss
functions that follow are only accessible in the neural network based
reconstruction framework. This unique property is owed to the
possibility that a neural network can be trained in advance on a
dataset of known input and output pairs. As pointed out in
\Cref{sec:specrecc}, a loss function can e.g.\ be defined directly on
a discretised representation of the spectral function $\rho$. This
approach is implemented through $L_\rho(\rho_\tinytext{sug})$, see
\labelcref{eq:Lrho}. The optimisation of the parameters
$\theta=\lbrace A_i, M_i, \Gamma_i \,|\, 0 \leq i <
N_{\textnormal{BW}}\rbrace$ of our chosen basis is an even more direct
approach. In principle, the space of all possible choices of
parameters is $\mathbb{R}_{+}^{3 \cdot N_{BW}}$, assuming they are all
positive definite. Of course, only finite subvolumes of this space
ever need to be considered as target spaces for reconstruction
methods. Therefore, we will often refer to a finite target volume
simply as the parameter space for a specific setting.  Accordingly, in
addition to the propagator and spectral function losses defined in
\Cref{eq:LGrho,eq:Lrho}, the respective parameter loss in this space
is given by:
\begin{align}
\label{eq:Theta}
L_\theta(\theta_\tinytext{sug}) = \lVert
\theta_\tinytext{sug} - \theta\rVert
\, .
\end{align}
All losses are evaluated using the 2-norm. In the case of the
parameter net, we have
$\rho_\tinytext{sug}\equiv\rho(\theta_\tinytext{sug})$. Apart from the
three given loss functions, we also investigate a combination of the
propagator and the spectral function loss,
\begin{align}
\label{eq:Combined}
  L_{G, \rho}(\rho_\tinytext{sug}, \alpha) = L_\rho(\rho_\tinytext{sug}) + \alpha L_G(\rho_\tinytext{sug})
  \,,
\end{align} %
where the parameter $\alpha$ determines the relative importance of the
two losses. In our experiments, we have chosen it such as to roughly
balance differences in the scales of the respective loss
functions. The type of loss function that is employed as well as the
selection of the training data have major impact on the resulting
performance of the neural network. Given this observation, it seems
likely that a further optimisation regarding the choice of the loss
function can significantly enhance the prediction quality. However,
for the time being, we content ourselves with the types given above
and postpone the exploration of more suitable training objectives to
future work.

\section{Numerical results} \label{sec:results}

In this section we present numerical results for the neural network
based reconstruction and validate the discussed potential advantages
by comparing to existing methods. Details on the training procedure as
well as the training and test datasets can be found in
\Cref{tab:volumes} and \Cref{sec:mocktst}, together with an
introduction to the used performance measures. We start now with a
brief summary of the main findings for our approach. Furthermore, a
detailed numerical analysis and discussion of different network setups
w.r.t\ performance differences are provided. Subsequently, additional
post-processing methods for an improvement of the neural network
predictions are covered. The section ends with a discussion of results
from the PoNet.  Readers who are interested in a comparison of the
neural network based reconstruction to existing methods may proceed
directly with \Cref{sec:benchmarking}.

\subsection{Reconstruction with neural networks}
\label{sec:reconneuralnetwork}

\begin{figure*}
	\includegraphics[width=0.48\textwidth]{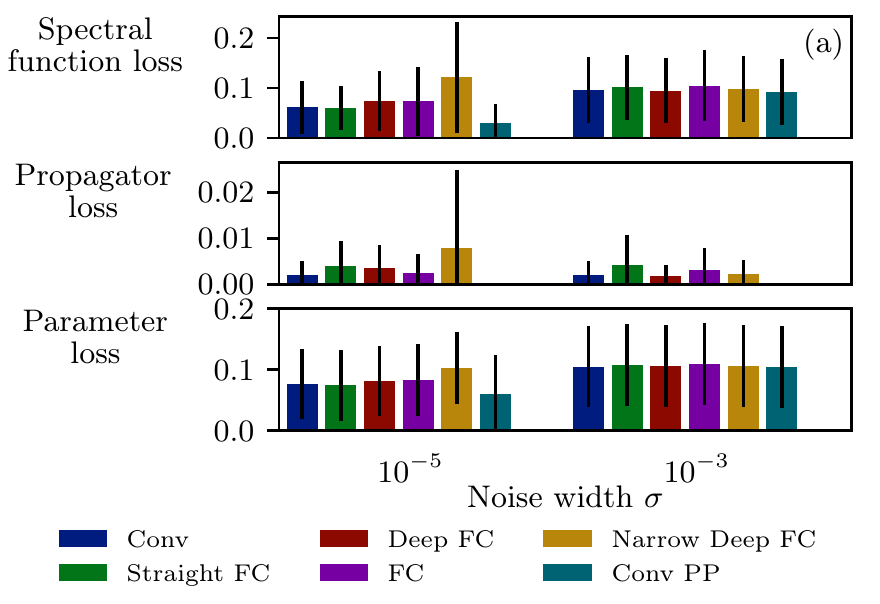}
	\hfill
	\includegraphics[width=0.48\textwidth]{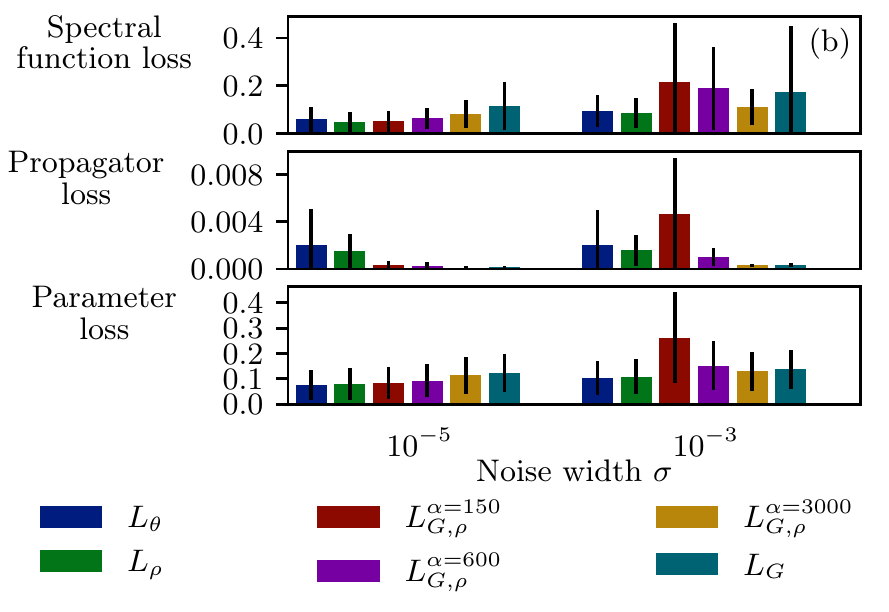}
	\caption{The performance of different net architectures and
          loss functions of a parameter net is compared for additive
          Gaussian noise with widths of $10^{-3}$ and $10^{-5}$ on the
          given input propagator: (a) Comparison of different net
          architecures. All networks are trained based on the
          parameter loss. The associated architectures can be found in
          \Cref{tab:netarchitecures}. (b) Comparison of different loss
          functions. Details on the loss functions are described at
          the end of \Cref{sec:trainingstrategy}. All results are
          based on networks with the architecture Conv. To (a) and
          (b): Shown here are the respective losses for the predicted
          parameters, for the discretised reconstructed spectral
          function and for the reconstructed propagator to the true,
          noise-free propagator. Both figures use the largest volume
          in parameter space, Vol O. The definitions of the
          performance measures are given at the end of
          \Cref{sec:mocktst}. The results on the left hand side imply
          that for larger errors, the choice of a specific network
          architecture has negligible impact on the quality of the
          reconstructions. All performance measures can be lowered for
          the given noise widths by applying a post-processing
          procedure on the suggested parameters of the network. In
          particular, the propagator loss can be minimised. The
          comparison on the right hand side shows that the choice of
          the loss function has major impact on the resulting
          performance of the network. The results underpin the
          importance of an appropriate loss function and support our
          argument of potential advantages of neural networks compared
          to existing approaches in \Cref{sec:specrecc}. Contour plots
          in parameter space are illustrated for the respective
          measures in \Cref{fig:contournetarchs} and
          \Cref{fig:contourloss}.}
	\label{fig:histonetarchs}
\end{figure*}

Our findings concerning the optimal setup of a feed-forward network
can be summarised as follows. As pointed out in \Cref{sec:specrecc},
the network aims to learn an approximate parametrisation of a manifold
of (matrix) inverses of the discretised K{\"a}ll{\'e}n-Lehmann
transformations. The inverse problem grows more severe if the
propagator values are afflicted with noise. In Bayesian terms, this is
caused by a wider posterior distribution for larger noise. The network
needs to have sufficient expressivity, i.e. an adequate number of
hyperparameters, to be able to learn a large enough set of inverse
transformations. We assume that for larger noise widths a smaller
number of hyperparameters is necessary to learn satisfactory
transformations, since the available information content about the
actual inverse transformation decreases for a respective exact
reconstruction. A varying severity of the inverse problem within the
parameter space leads to an optimisation of the spectral
reconstruction in regions where the problem is less severe. This
effect occurs naturally, since there the network can minimise the loss
more easily than in regions where the problem is more severe. Besides
the severity of the inverse problem, the form of the loss function has
a large impact on global optima within the landscape of the solution
space. Based on these observations, an appropriate training of the
network is non-trivial and demands a careful numerical analysis of the
inverse problem itself, and of different setups of the optimisation
procedure. A sensible definition of the loss function or a non-uniform
selection of the training data are possible approaches to address the
disparity in the severity of the inverse problem. A more
straightforward approach is to iteratively reduce the covered
parameter ranges within the learning process, based on previous
suggested outcomes. This amounts to successively increasing the
prediction accuracy by restricting the network to smaller and smaller
subvolumes of the original solution space. However, one should be
aware that this approach is only sensible if the reconstructions for
different noise samples on the original propagator data are
sufficiently close to each other in the solution space. A successive
optimisation of the prediction accuracy in such a way can also be
applied to existing methods. All approaches ultimately aim at a more
homogeneous reconstruction loss within the solution space. This allows
for a reliable control of systematic errors, as well as an accurate
estimation of statistical errors. The desired outcome for a generic
set of Breit-Wigner parameters is illustrated and discussed in
\Cref{fig:numbwimpact}.

The quality and reliability of reconstructions heavily depend on the
following details of the training procedure and the inverse problem
itself, with varying levels of impact given a specific situation:

\begin{itemize}
	\item local differences in the severity of the inverse problem
	\item information loss in the forward pass and due to
          statistical noise
	\item loss function / prior information
	\item complexity / expressivity of the network architecture
\end{itemize}
In essence, we wish to emphasise that a reliable reconstruction is a
multifactorial problem whose facets need to be disentangled in order
to understand all contributions to systematic errors.

\begin{figure*}
	\includegraphics[width=0.193\textwidth]{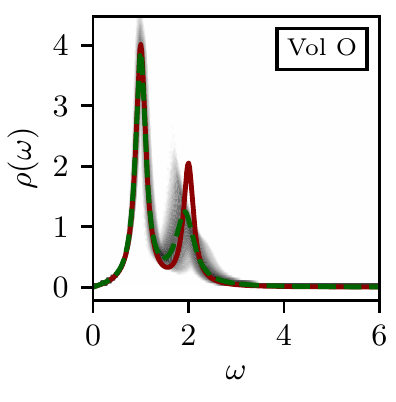} \includegraphics[width=0.1556\textwidth]{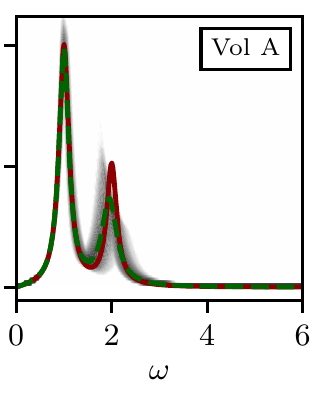} \includegraphics[width=0.1556\textwidth]{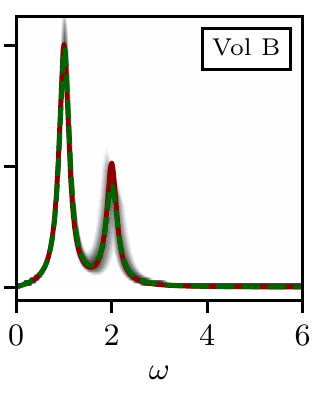} \includegraphics[width=0.1556\textwidth]{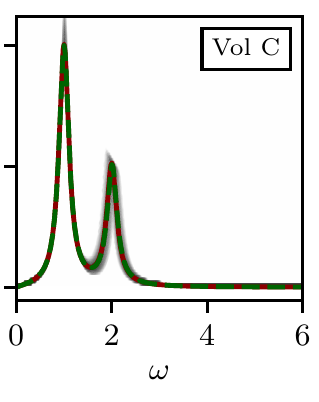} \includegraphics[width=0.1556\textwidth]{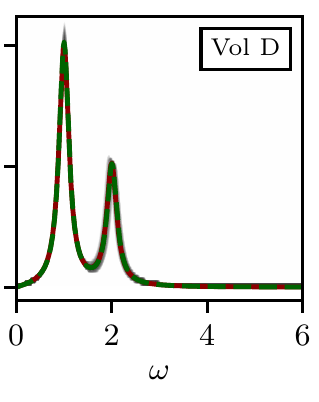} \includegraphics[width=0.149\textwidth]{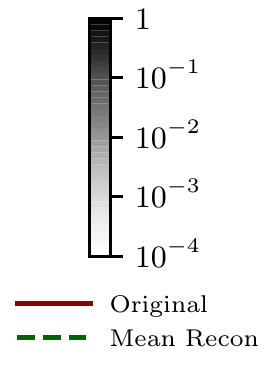}
	\caption{The uncertainties of reconstructions of spectral
          functions on the same original propagator are illustrated in
          the same manner as described in \Cref{fig:numbwimpact} for
          different volumes of the parameter space, again using a
          noise width of $10^{-3}$. The plots demonstrate how the
          quality of the reconstruction improves if the parameter
          space which the network has to learn is decreased. The
          volumes of the corresponding parameter spaces are listed in
          \Cref{tab:volumes}. The results are computed from the Conv
          PaNet. The systematic deviation of the distribution of
          reconstructions for large volumes shows that the network has
          not captured the manifold of inverse transformations
          completely for the entire parameter space.  This is in
          concordance with the results discussed in
          \Cref{fig:contournetarchs} and \Cref{fig:contourvol}.}
	\label{fig:volumeimpact}
\end{figure*}

\begin{figure*}
	\includegraphics[width=1\textwidth]{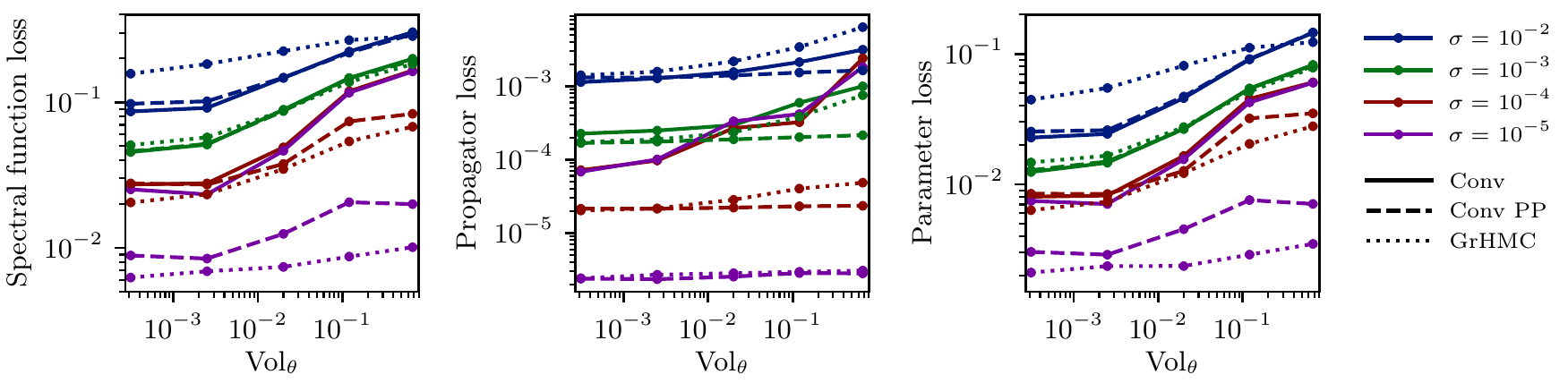}
	\caption{The plots in this figure quantify the impact of the
          parameter space volume used for the training on the
          performance of the parameter network. The performance
          measures are computed based on the test set of the smallest
          volume, Vol D. The parameter ranges in the training set are
          gradually reduced to analyse different levels of complexity
          of the problem. Separate networks are trained for each
          volume, which are listed in \Cref{tab:volumes}. The results
          demonstrate the potential advantage of an iterative
          restriction of the parameter ranges of possible
          solutions. The contour plots in \Cref{fig:contourvol} depict
          changes of the performance measures within the parameter
          space. More strongly peaked prior distributions lead to
          better reconstructions. The comparison with results of the
          GrHMC approach illustrates the improvement of the
          performance of neural networks for larger errors and smaller
          volumes. These observations confirm the discussions of
          \Cref{fig:volumeimpact} and \Cref{fig:errorimpact}. Adding a
          post-processing step leads in particular for the propagator
          loss and for smaller noise widths to an improvement of the
          reconstruction, as has also been discussed in
          \Cref{fig:histonetarchs}.} \label{fig:comparisonvolume}
\end{figure*}

\begin{figure*}
	\includegraphics[width=1\textwidth]{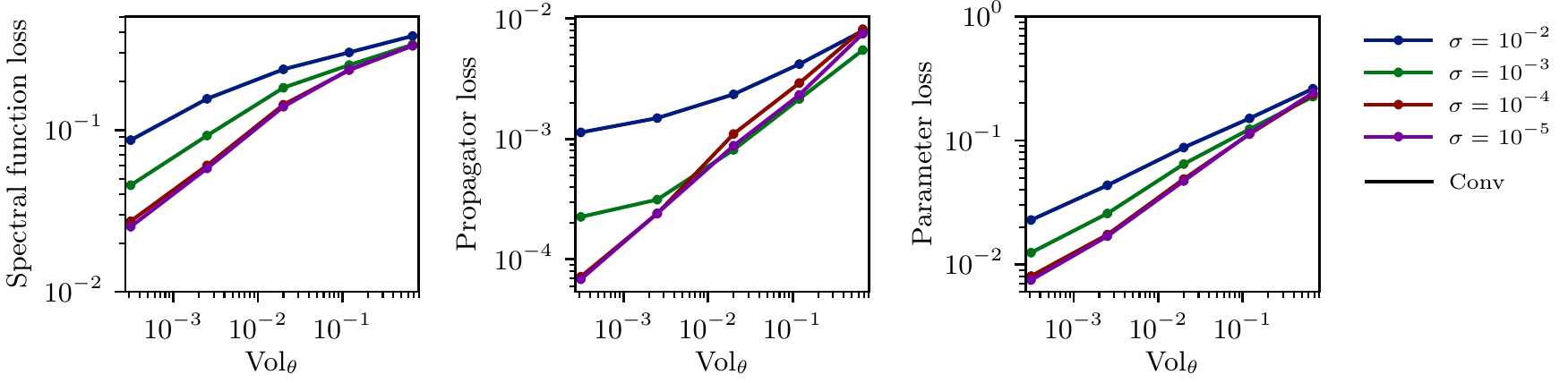}
	\caption{Comparison of reconstruction errors of the Conv PaNet
          trained only on the smallest Vol D for different noise
          levels, evaluated with test volumes which are also larger
          than D. The test datasets are equivalent to the ones used
          for the other tasks described in this paper. In contrast to
          \Cref{fig:comparisonvolume}, which shows the prediction
          quality as a function of the training volume with a fixed
          test volume Vol D, here the performance is evaluated as a
          function of the test volume using a fixed training
          volume.} \label{fig:comparisonoutofrange}
\end{figure*}

\begin{figure*}
	\includegraphics[width=0.78\textwidth]{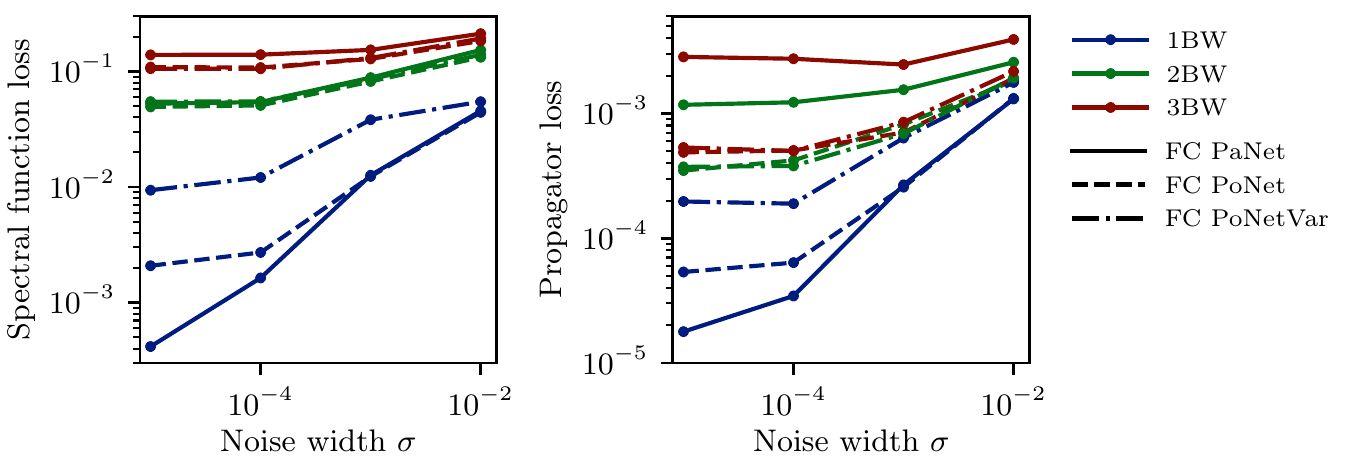}
	\caption{Comparison of reconstruction errors of the PaNet and
          the PoNet for the FC architecture.  The performance measures
          are computed based on the test set of the largest parameter
          space volume Vol O for one, two and three Breit-Wigners. All
          networks are trained based on the parameter ranges of Vol
          O. Loss functions are the parameter loss $L_{\theta}$ for
          the PaNet and the spectral function loss $L_{\rho}$ for both
          PoNets. The overall smaller losses for the point nets are
          due to the large number of degrees of freedom for the
          point-like representation of the spectral function. The
          partly competitive performance of the PoNetVar compared to
          the results of the PoNet encourage the further investigation
          of networks that are trained using a more exhaustive set of
          basis functions to describe physical structures in the
          spectral functions.}
	\label{fig:comppoint}
\end{figure*}

The impact of the net architecture and the loss function on the
overall performance within the parameter space is illustrated in
\Cref{fig:histonetarchs}. Associated contour plots can be found in the
appendix, \Cref{fig:contournetarchs,fig:contourloss}. These plots
demonstrate that the minima in the loss landscape highly depend on the
employed loss function. In turn, this leads to different performance
measures. This observation confirms our previous discussion and the
necessity of an appropriate definition of the loss function. It also
reinforces our arguments regarding potential advantages of neural
networks in comparison to other approaches for spectral
reconstruction. The comparison of different feed-forward network
architectures shows that the specific details of the network structure
are rather irrelevant, provided that the expressivity is sufficient.

\begin{figure*}
	\includegraphics[width=0.236\textwidth]{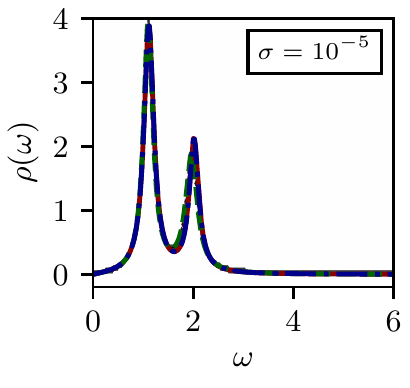} \includegraphics[width=0.192\textwidth]{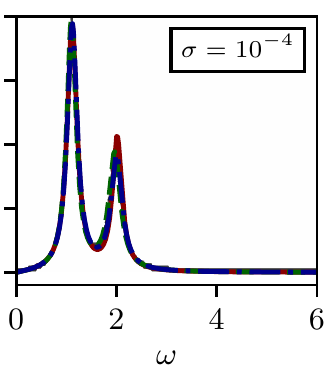} \includegraphics[width=0.192\textwidth]{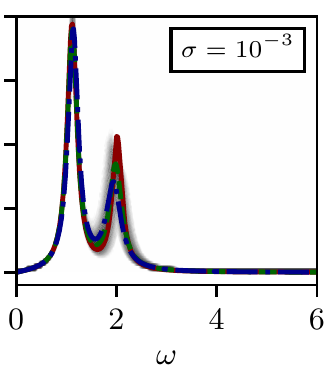} \includegraphics[width=0.192\textwidth]{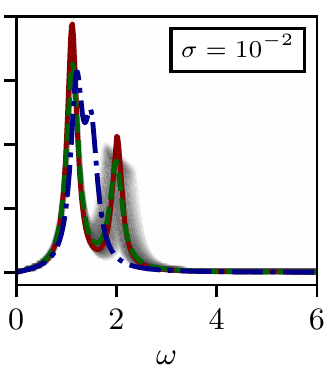} \includegraphics[width=0.13\textwidth]{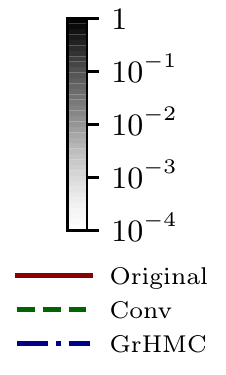}
	\caption{The quality of the reconstruction of two Breit-Wigner
          peaks is compared for different strength of additive noise
          on the same propagator. The labels indicate the noise width
          on the original propagator. It can be seen that the
          reconstructed spectral function of the neural network
          exhibits in particular for larger errors a lower deviation
          to the original spectral function than the GrHMC
          method. This mirrors the in general observable better
          performance of the neural network for larger errors, as can
          be seen in \Cref{fig:comparisonvolume} and in
          \Cref{fig:comparisonall}. The green and the red curve
          correspond to reconstructions of the Conv PaNet and the
          GrHMC method for the same given noisy propagator. The prior
          is in both cases given by the parameter range of volume Vol
          B. The uncertainty of the reconstructions for the neural
          network is depicted by the grey shaded areas as described in
          \Cref{fig:numbwimpact}. For small errors, this area is
          covered by the corresponding reconstructed spectral
          functions.}
	\label{fig:errorimpact}
\end{figure*}

\begin{figure*}
	\includegraphics[width=1\textwidth]{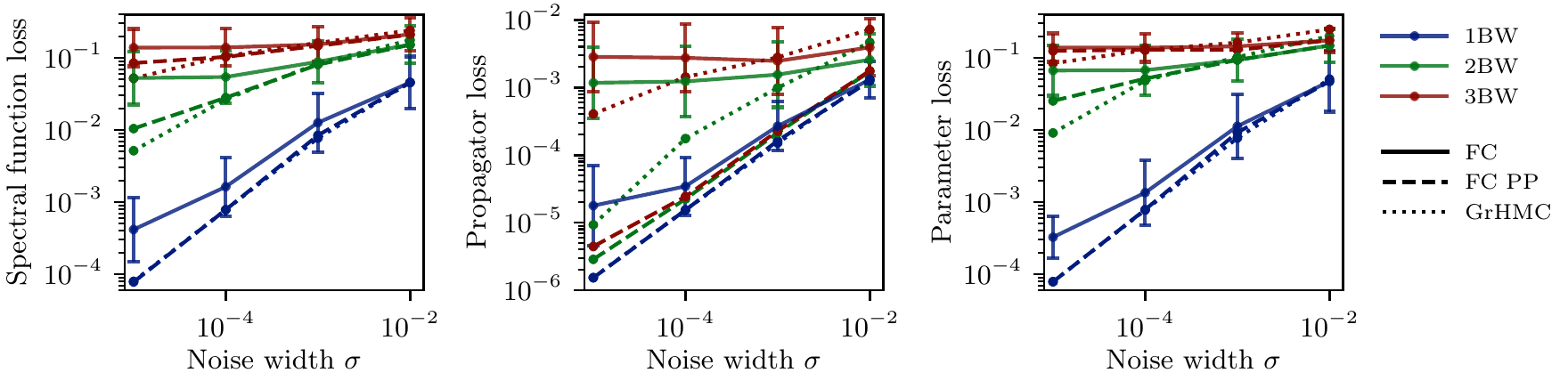}
	\caption{The performance of the reconstruction of spectral
          functions is benchmarked for the parameter network, which is
          trained with $L_{\theta}$, by a comparison to results of the
          GrHMC method. The parameter network is in particular for
          large noise widths competitive. The worse performance for
          smaller noise widths is a result of an inappropriate
          training procedure and a too low expressive power of the
          neural network. The problems are caused by a varying
          severity of the inverse problem and by a too large parameter
          space that needs to be covered by the neural network, as
          discussed in \Cref{sec:reconneuralnetwork}. The error bars
          of the results for the FC network are representative for
          typical errors within all methods and plots of this kind.}
	\label{fig:comparisonall}
\end{figure*}

Differences in the performance of the networks that are trained with
the same loss function become less visible for larger noise. This is
illustrated by a comparison of contour plots with different noise
widths, see e.g.\ \Cref{fig:contournetarchs}. The severity of the
inverse problem grows with the noise and the information content about
the actual matrix transformation decreases. These properties lead to
the observation of a generally worse performance for larger noise
widths, as can be inferred from \Cref{fig:volumeimpact}, as well as
\Cref{fig:errorimpact,fig:comparisonall}, which are discussed
later. They also imply that for specific noise widths, the neural
network possesses enough hyperparameters to learn a sufficient
parametrisation of the inverse transformation manifold. Furthermore,
the local optima into which the network converges are mainly
determined by differences in the local severity of the inverse
problem. Hence, the issue remains that generic loss functions are
inappropriate to address the varying local severity of the inverse
problem. This issue implies the existence of systematic errors for
particular regions within the parameter space, as can be seen e.g.\ in
the left plot of \Cref{fig:volumeimpact}.

\begin{figure}
	\includegraphics[width=1\textwidth]{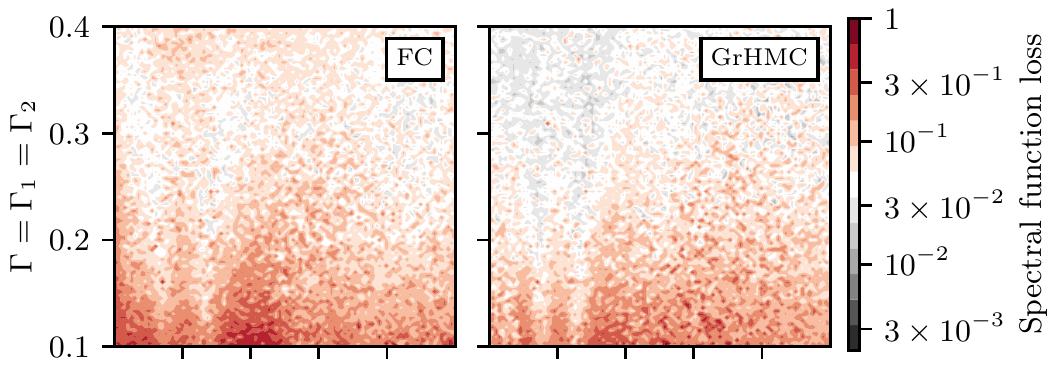}\\
	\includegraphics[width=1\textwidth]{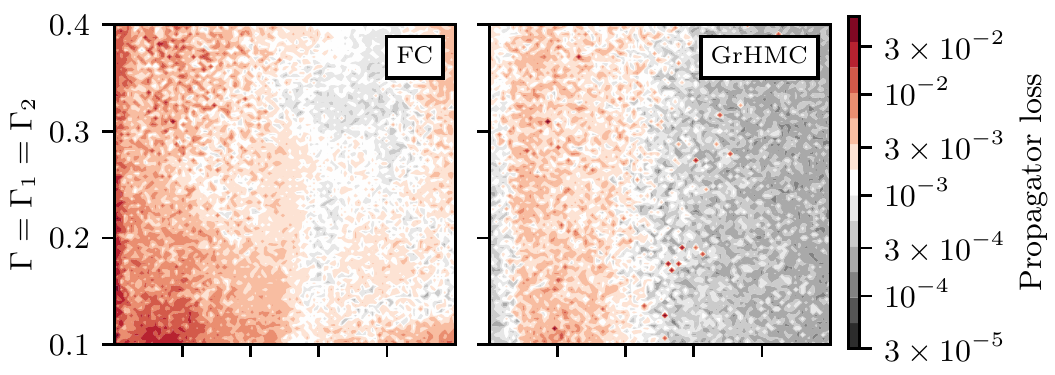}\\
	\includegraphics[width=1\textwidth]{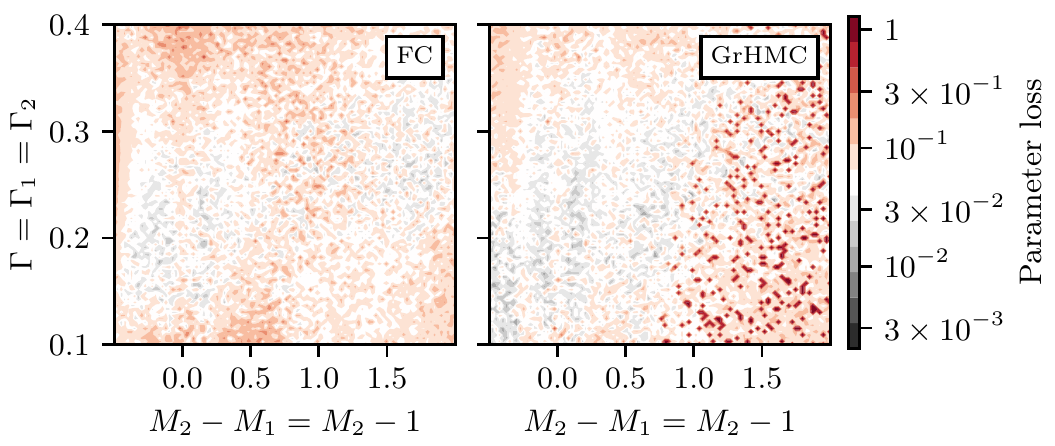}
	\caption{Comparison of performance measures for the
          reconstruction of two Breit-Wigners of the FC parameter
          network with the GrHMC method for input propagators with
          noise width $10^{-3}$ within the parameter space volume Vol
          O. The similar loss landscape emphasises the high impact of
          variations of the severity of the inverse problem within the
          parameter space on the quality of reconstructions. Contrary
          to expectations, the parameter network mimics, despite an
          optimisation based on the parameter loss $L_\theta$, the
          reconstruction of the GrHMC method which relies on an
          optimisation of the propagator loss $L_G$ with respect to
          the parameters. A reconstruction resulting in an averaged
          peak with the other parameter set effectively removed, as
          outlined in \cite{Cyrol:2018xeq}, results in the spiking
          parameter losses for the GrHMC reconstructions with large
          errors.}
	\label{fig:comparisoncontourall}
\end{figure}

\begin{table}
	\begin{tabular}{cccccc}
		Vol & A & M & $\Gamma$ & $\Delta M$\\
		\hline\hline
		O & $[0.1, 1.0]$ & $[0.5, 3.0]$ & $[0.1, 0.4]$ & $[0.0, 2.5]$\\
		A & $[0.3, 0.7]$ & $[0.5, 3.0]$ & $[0.1, 0.3]$ & $[0.25, 1.75]$\\
		B & $[0.4, 0.6]$ & $[0.5, 3.0]$ & $[0.1, 0.2]$ & $[0.5, 1.5]$\\
		C & $[0.45, 0.55]$ & $[0.5, 3.0]$ & $[0.1, 0.15]$ & $[0.75, 1.25]$\\
		D & $[0.475, 0.525]$ & $[0.5, 3.0]$ & $[0.1125, 0.1375]$ & $[0.875, 1.125]$\\
	\end{tabular}
	\caption{Parameter ranges of the different volumes in
          parameter space used for training. Parameters are sampled
          uniformly based on the given bounds for the training and
          test sets. For the case of two and three Breit-Wigner
          functions, the difference in mass $\Delta M = M_2-M_1$ is
          limited to restrict the minimum possible distance between
          two peaks. The volumes $V_\theta$ are computed based on
          these parameter ranges.}
	\label{tab:volumes}
\end{table}

The results shown in \Cref{fig:volumeimpact,fig:comparisonvolume} as
well as \Cref{fig:contourvol} in the appendix confirm our discussion
regarding the expressive power of the network w.r.t.\ the complexity
of the solution space and the decreasing information content for
larger errors. The parameter space is gradually reduced, effectively
increasing the expressivity of the network relative to the severity of
the problem and improving the behavior of the loss function for a
given fixed parameter space. The respective volumes are listed in
\Cref{tab:volumes}. Shrinking the parameter space leads to a more
homogeneous loss landscape due to the increased locality, thereby
mitigating the issue of inappropriate loss functions. The necessary
number of hyperparameters decreases for larger noise widths and
smaller parameter ranges in the training and test dataset. The
arguments above imply a better performance of the network for smaller
parameter spaces. A reduction of the parameter space effectively
corresponds to a sharpening of the prior information, which also has
positive effects on the spread of the posterior distribution. More
detailed discussions on the impact of different elements of the
training procedure can be found in the captions of the respective
figures.

Since increasing the expressivity of the network is limited by the
computational demand required for the training, one can also apply
post-processing methods to improve the suggested outcome w.r.t.\ the
initially given, noisy propagator. These methods are motivated by the
in some cases large observed root-mean-square deviation of the
reconstructed suggested propagator to the input, see for example again
\Cref{fig:histonetarchs}. The application of standard optimisation
methods on the suggested results of the network represents one
possible approach to address this problem. Here, the network's
prediction is interpreted as a guess of the MAP estimate, which is
presumed to be close to the true solution. For the PaNet, we minimise
the propagator loss a posteriori with respect to the following loss
function:
\begin{align}
  \min_{\theta_\tinytext{sug}}\;L_{\textnormal{PP}}[\theta_\tinytext{sug}]=\min_{\theta_\tinytext{sug}}\;\lVert
  G_\textnormal{noisy} -
  G\left[\rho\left(\theta_\tinytext{sug}\right)\right] \rVert\,.
\end{align}
This ensures that suggestions for the reconstructed spectral functions
are in concordance with the given input propagator. Results obtained
with an additional post-processing are marked by the attachment PP in
this work. The numerical results in
\Cref{fig:volumeimpact,fig:comparisonall} show that the finite size of
the neural network can be partially compensated for small errors. The
resulting low propagator losses are noteworthy, and are close to
state-of-the-art spectral reconstruction approaches. One reason for
this similarity is the shared underlying objective function. However,
the situation is different for larger noise widths. For our choice of
hyperparameters, the algorithm quickly converges into a local minimum.
For large noise widths, the optimisation procedure may even lead to
worse results than the initially suggested reconstruction. This is due
to the already mentioned systematic deviations which are caused by the
inappropriate choice of the loss function for large parameter spaces.
This kind of post-processing should therefore be applied with caution,
since it may cancel out the potential advantages of neural networks
w.r.t.\ the freedom in the definition of the loss function.

The following alternative post-processing approach preserves the
potential advantages of neural networks while nevertheless minimising
the propagator loss. The idea is to include the network into the
optimisation process through the following objective:
\begin{align} \min_{G_\textnormal{input}}\;L_{\textnormal{input}}[G_\textnormal{input}]=\min_{G_\textnormal{input}}\;\lVert G_\textnormal{noisy} -
G\left[\rho\left(\theta_\tinytext{sug}\right)\right] \rVert\,,
\end{align} %
where $G_\textnormal{input}$ corresponds to the input propagator of
the neural network and $\theta_\tinytext{sug}$ to the associated
outcome. This facilitates a compensation of badly distributed noise
samples and allows a more accurate error estimation. The approach is
only sensible if no systematic errors exist for reconstructions within
the parameter space, and if the network's suggestions are already
somewhat reliable. We postpone a numerical analysis of this
optimisation method together with the exploration of more appropriate
loss functions and improved training strategies to future work, due to
a currently lacking setup to train such a network.

\begin{figure*}
	\includegraphics[width=\textwidth]{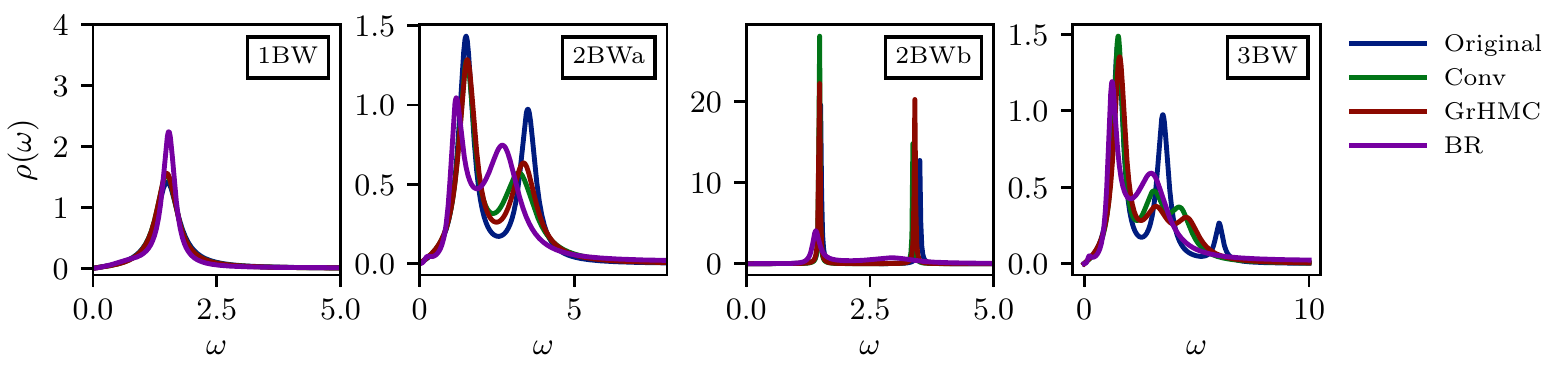}
	\caption{Reconstructions of one, two and three Breit-Wigners
          are compared for our proposed neural network approach, the
          GrHMC method and the BR method. The reconstructions of the
          first two methods are based on a single sample with noise
          width $10^{-3}$, while the results of the BR method are
          obtained from multiple samples with larger errors, but an
          average noise width of $10^{-3}$ as well. In contrast to the
          previous plots, the neural network and the GrHMC method now
          use different priors for each case in order to allow for a
          reasonable comparison with the BR method,
          see~\Cref{tab:volumesbrcompairson}. We observe that all
          approaches qualitatively capture the features in the
          spectral function. Due to the comparably large error on the
          input data, all methods are expected to face difficulties in
          finding an accurate solution. The reconstructions of the
          neural network approach and the GrHMC method are comparable,
          whereas the BR method struggles in particular with thin
          peaks and the three Breit-Wigner case. The results
          demonstrate that, generally, using suitable basis functions
          and incorporating prior information lead to a superior
          reconstruction performance.}
	\label{fig:brcomparison}
\end{figure*}

\Cref{fig:comparisonoutofrange,fig:outofparameterspace} serve to
quantify the generalization capability of our neural network approach
with regard to data that lie outside of the training
volume. \Cref{fig:comparisonoutofrange} shows a comparison of error
metrics obtained with the Conv PaNet trained only on the smallest Vol
D when applied to larger parameter volumes, at a few different noise
levels. As we expect, the performance decreases with larger test
volumes, but loss values notably remain in the same orders of
magnitude that are observed for larger training volumes, indicating
that the generalization capability of the network has at first only a
rather weak dependence on the ratio between test and training volume,
which only grows more severe if the test volume becomes much
larger. This is further illustrated by \Cref{fig:outofparameterspace},
showing a rather flat loss landscape in the immediate vicinity of the
training volume boundaries without sharp transitions, and gradual
worsening of the prediction quality as one moves further away. We
conclude that our approach is moderately robust against deviations of
the true solution from the considered training volume and only fails
at larger distances.

In \Cref{fig:comppoint,fig:contourpoint}, results of the PoNet and the
PaNet are compared. We observe that spectral reconstructions based on
the PoNet structure suffer from similar problems as the PaNet. The
point-like representation of the spectral function introduces a large
number of degrees of freedom for the solution space. The training
procedure implicitly regularises this problem, however, a visual
analysis of individual reconstructions shows that in some cases the
network struggles with common issues known from existing methods, such
as partly non-Breit-Wigner like structures and wiggles. An application
of the proposed post-processing methods serves as a possible approach
to circumvent such problems. An inclusion of further regulator terms
into the loss function, concerning e.g.\ the smoothness of the
reconstruction, is also possible.

\subsection{Benchmarking and discussion} \label{sec:benchmarking}

In this section, we want to emphasise differences of our proposed
neural network approach to existing methods. Our arguments are
supported by an in-depth numerical comparison.

Within all approaches the aim is to map out, or at least to find the
maximum of, the posterior distribution $P(\rho|G)$ for a given noisy
propagator $G$. The BR and GrHMC methods represent iterative
approaches to accomplish this goal. The algorithms are designed to
find the maximum for each propagator on a case-by-case basis. The
GrHMC method additionally provides the possibility to implement
constraints on the functional basis of the spectral function in a
straightforward manner. In contrast, a neural network aims to learn
the full manifold of inverse K\"allen-Lehman transformations for any
noisy propagator (at least within the chosen parameter space). In this
sense, it needs to propose for each given propagator an estimate of
the maximum of $P(\rho|D)$. A complex parametrisation, as given by the
network, an exhaustive training dataset and the optimisation procedure
itself are essential features of this approach for tackling this tough
challenge. The computational effort to find a solution in an iterative
approach is therefore shifted to the training process as well as the
memory demand of the network. Accordingly, the neural network based
reconstruction can be performed much faster after training has been
completed, which is in particular advantageous when large sets of
input propagators are considered. In our experiments, the time
required for running the GrHMC algorithm and training the networks was
roughly similar, generally being of the order of a few hours. A
quantitative comparison is difficult due to the use of completely
distinct software packages and different utilisation of hardware
accelerators. However, we emphasise again that one run of the GrHMC
can only provide a prediction for one specific propagator, whereas the
trained network can be evaluated quickly on large datasets and
additionally allows fast retraining when different data are expected,
without having to start from scratch.

The numerical results in
\Cref{fig:comparisonvolume,fig:errorimpact,fig:comparisonall,fig:comparisoncontourall,fig:brcomparison}
demonstrate that the formal arguments of \Cref{sec:specrecc} apply,
particularly for comparably large noise widths as well as smaller
parameter ranges. For both cases, the network successfully
approximates the required inverse transformation manifold. Smaller
noise widths and a larger set of possible spectral functions can be
addressed by increasing the number of hyperparameters and through the
exploration of more appropriate loss functions, as was already
discussed previously.

\section{Conclusion} \label{sec:conclusion}

In this study we have explored artificial neural networks as a tool to
deal with the ill-conditioned inverse problem of reconstructing
spectral functions from noisy Euclidean propagator data. We
systematically investigated the performance of this approach on
physically motivated mock data and compared our results with existing
methods. Our findings demonstrate the importance of understanding the
implications of the inverse problem itself on the optimisation
procedure as well as on the resulting predictions.

The crucial advantage of the presented ansatz is the superior
flexibility in the choice of the objective function. As a result, it
can outperform state-of-the-art methods if the network is trained
appropriately and exhibits sufficient expressivity to approximate the
inverse transformation manifold. The numerical results demonstrate
that defining an appropriate loss function grows increasingly
important for an increased variability of considered spectral
functions and of the severity of the inverse problem.

In future work, we aim to further exploit the advantage of neural
networks that local variations in the severity of the inverse problem
can be systematically compensated. The goal is to eliminate systematic
errors in the predictions in order to facilitate a reliable
reconstruction with an accurate error estimation. This can be realised
by finding more appropriate loss functions with the help of implicit
and explicit approaches~\cite{Santos2017,Wu2018}. A utilisation of
these loss functions in existing methods is also possible if they are
directly accessible. Varying the prior distribution will also be
investigated, by sampling non-uniformly over the parameter space
during the creation of the training data. Furthermore, we aim at a
better understanding of the posterior distribution through the
application of invertible neural networks~\cite{Rother2018}. This
novel architecture provides a reliable estimation of errors by mapping
out the entire posterior distribution by construction.

In conclusion, we believe that the suggested improvements will boost
the performance of the proposed method to an as of yet unprecedented
level and that neural networks will eventually replace existing
state-of-the-art methods for spectral reconstruction.

%%%%%%%%%%%%%%%%%%%%%%%%%%%%%%%%%%%%%%%%%%%%%%%%%%%%%%%%%%%

\begin{acknowledgments}
  We thank S.~Bl\"ucher, A.~Cyrol and I.-O.~Stamatescu for
  discussions. M. Scherzer acknowledges financial support from DFG
  under STA 283/16-2. F.P.G.~Ziegler acknowledges support from
  Heidelberg University. This work is supported by the Deutsche
  Forschungsgemeinschaft (DFG, German Research Foundation) under
  Germany's Excellence Strategy EXC 2181/1 - 390900948 (the Heidelberg
  STRUCTURES Excellence Cluster) and under the Collaborative Research
  Centre SFB 1225 (ISOQUANT), the ExtreMe Matter Institute and the
  BMBF Grant No. 05P18VHFCA.
\end{acknowledgments}

\appendix

\section{BR method}
\label{app-sec:BR}

Different Bayesian methods propose different prior probabilities,
i.e.\ they encode different types of prior information. The well-known
Maximum Entropy Method e.g.\ features the Shannon-Jaynes entropy
\begin{align}
S_{\rm SJ}=\int d\omega \big( \rho(\omega)- m(\omega) -
\rho(\omega){\rm log}\big[ \frac{\rho(\omega)}{m(\omega)} \big]\big),
\end{align}
while the more recent BR method uses a variant of the gamma
distribution
\begin{align}
S_{\rm BR}=\int d\omega \big( 1-
\frac{\rho(\omega)}{m(\omega)} + {\rm log}\big[
\frac{\rho(\omega)}{m(\omega)} \big]\big).
\end{align}
Both methods e.g.\ encode the fact that physical spectral functions
are necessarily positive definite but are otherwise based on different
assumptions.

As Bayesian methods they have in common that the prior information has
to be encoded in the functional form of the regulator and the supplied
default model $m(\omega)$. Note that discretising $\rho$ by choosing a
particular functional basis also introduces a selection of possible
outcomes. The dependence of the most probable spectral function, given
input data and prior information, on the choice of $S$, $m(\omega)$
and the discretised basis comprises the systematic uncertainty of the
method.

One major limitation to Bayesian approaches is the need to formulate
our prior knowledge in the form of an optimisation functional. The
reason is that while many of the correlation functions relevant in
theoretical physics have very well defined analytic properties it has
not been possible to formulate these as a closed regulator functional
$S$. Take as an example the retarded propagator (for a more
comprehensive discussion see \cite{Cyrol:2018xeq}). Its analytic
structure in the imaginary frequency plane splits into two parts, an
analytic half-plane, where the Euclidean input data is located, and a
meromorphic half-plane which contains all structures contributing to
the real-time dynamics. Encoding this information in an appropriate
regulator functional has not yet been achieved.

Instead the MEM and the BR method rather use concepts unspecific to
the analytic structure, such as smoothness, to derive their
regulators. Among others this e.g.\ manifests itself in the presence
of artificial ringing, which is related to unphysical poles
contributing to the real-time propagator, which however should be
suppressed by a regulator functional aware of the physical analytic
properties.

\section{GrHMC method}

The main idea of the setup is already stated in the main text in
\Cref{sec:theory} and was first introduced in \cite{Cyrol:2018xeq}.
Nevertheless, for completeness we outline the entire reconstruction
process here. The approach is based on formulating the basis expansion
in terms of the retarded propagator. The resulting set of basis
coefficients are then determined via Bayesian inference. This leaves
us with two objects to specify in the reconstruction process, the
choice of a basis/ansatz for the retarded propagator and suitable
priors for the inference.

Once a basis has been chosen it is straightforward to write down the
corresponding regression model. As in the reconstruction with neural
nets we use a fixed number of Breit-Wigner structures, c.f.
\labelcref{eq:breitwigner}, corresponding to simple poles in the
analytically continued retarded propagator. The logarithm of all
parameters is used in the model in order to enforce positivity of all
parameters. The uniqueness of the parameters is ensured by using an
ordered representation of the logarithmic mass parameters.

The other crucial point is the choice of priors, which are of great
importance to tame the ill-conditioning practically and should
therefore be chosen as restrictive as possible. For comparability to
the neural net reconstruction, the priors are matched to the training
volume in parameter space. However, it is more convenient to work with
a continuous distribution. Hence the priors of the logarithmic
parameters are chosen as normal distributions where we have fixed the
parameters by the condition that the mean of the distribution is the
mean of the training volume and the probability at the boundaries of
the trainings volume is equal. Details on the training volume in
parameter space can be found in \Cref{tab:volumes} and
\Cref{sec:mocktst}.

All calculations for the GrHMC method are carried out using the python
interface \cite{PyStan2018} of Stan \cite{carpenter2017stan}.

\section{Mock data, training set and training procedure} \label{sec:mocktst}

We consider three different levels of difficulty for the
reconstruction of spectral functions to analyse and compare the
performance of the approaches in this work. These levels differ by the
number of Breit-Wigners that need to be extracted based on the given
information of the propagator. We distinguish between training and
test sets with one, two and three Breit-Wigners. A variable number of
Breit-Wigners within a test set entails the task to determine the
correct number of present structures. This can be done a priori or a
posteriori based either on the propagator or on the quality of the
reconstruction. While it is straightforward to implement this for the
PoNet, it is not completely clear how one should proceed for the
PaNet. We postpone this problem to future work.

The training set is constructed by sampling parameters uniformly
within a given range for each parameter. The ranges for the parameters
of a Breit-Wigner function of \labelcref{eq:breitwigner} are as
follows: $M\in[0.5, 3.0]$, $\Gamma\in[0.1, 0.4]$ and $A\in[0.1, 1.0]$.
In addition, we investigate the impact of the size of the parameter
space on the performance of the network for the case of two
Breit-Wigner functions. This is done by decreasing the ranges of the
parameters $\Gamma$ and $A$ gradually. We proceed differently for the
two masses to guarantee a certain finite distance between the two
Breit-Wigner peaks. Instead of decreasing the mass range, the minimum
and maximum distance of the peaks is restricted. Details on the
different parameter spaces were stated in \Cref{tab:volumes}. The
propagator function is parametrised by $N_p=100$ data points that are
evaluated on a uniform grid within the interval $\omega\in[0, 10]$.
For a training of the point net, the spectral function is discretised
by $N_{\omega}=500$ data points on the same interval. Details about
the training procedure can be found at the end of the section. The
parameter ranges deviate for the comparison of the neural network
approach with existing methods. The corresponding ranges are listed in
\Cref{tab:volumesbrcompairson}. To avoid any confusion,
\Cref{tab:networktrainingdetails} provides a comprehensive list of all
figures with the associated model details and parameter ranges.

\begin{table}
	\begin{tabular}{ccccc}
		BR Comparison & A & M & $\Gamma$ \\
		\hline\hline
		1BW & $[0.1, 1.0]$ & $[0.5, 3.0]$ & $[0.1, 0.4]$ \\
		2BWa & $[0.2, 1.8]$ & $[0.8, 3.8]$ & $[0.2, 1.0]$ \\
		2BWb & $[0.3, 1.2]$ & $[0.8, 3.8]$ & $[0.002, 0.02]$\\
		3BW & $[0.2, 1.8]$ & $[1.0, 6.0]$ & $[0.2, 1.0]$\\
	\end{tabular}
	\caption{Parameter ranges for the training of the neural networks for the comparison in~\Cref{fig:brcomparison}.}
	\label{tab:volumesbrcompairson}
\end{table}

The different approaches are compared by a test set for each number of
Breit-Wigners consisting of $1000$ random samples within the parameter
space. Another test set is constructed for two Breit-Wigners with a
fixed scaling $A_1=A_2=0.5$, a fixed mass $M_1=1$ and equally chosen
widths $\Gamma:=\Gamma_1=\Gamma_2$. The mass $M_2$ and the width
$\Gamma$ are varied according to a regular grid in parameter space.
This test set allows the analysis of contour plots of different loss
measures. It provides more insights into the minima of the loss
functions of the trained networks and into the severity of the inverse
problem. The contour plots are averaged over $10$ samples for the
noise width of $10^{-3}$ (except for \Cref{fig:comparisoncontourall}).

We investigate three different performance measures and different
setups of the neural network for a comparison to existing methods. The
root-mean-square-deviation of the predicted parameters in parameter
space, of the reconstructed spectral function and of the reconstructed
propagator are considered. For the latter case, the error is computed
based on the original propagator without noise. The spectral function
loss and the propagator loss are computed based on the discretised
representations on the uniform grid. Representative error bars for all
methods are depicted in \Cref{fig:comparisonall}.

The training procedure for the neural networks in this work is as
follows. A separate neural network is trained for each training set,
i.e., for each error and for each range of parameters. The learning
rates are between $10^{-5}$ and $10^{-7}$. The batch size is between
$128$ and $500$ and the number of generated training samples per epoch
is around $6 \times 10^{5}$. Depending on the kind of network, the
nets are trained for $80$ to $160$ epochs. The used loss functions are
described at the end of \Cref{sec:trainingstrategy}. The implemented
net architectures are provided in \Cref{tab:netarchitecures}. Details
about the training of the different networks and about the respective
utilized test set for the evaluation can be found in
\Cref{tab:networktrainingdetails} for each figure.

\begin{figure*}
  \includegraphics[width=1\textwidth]{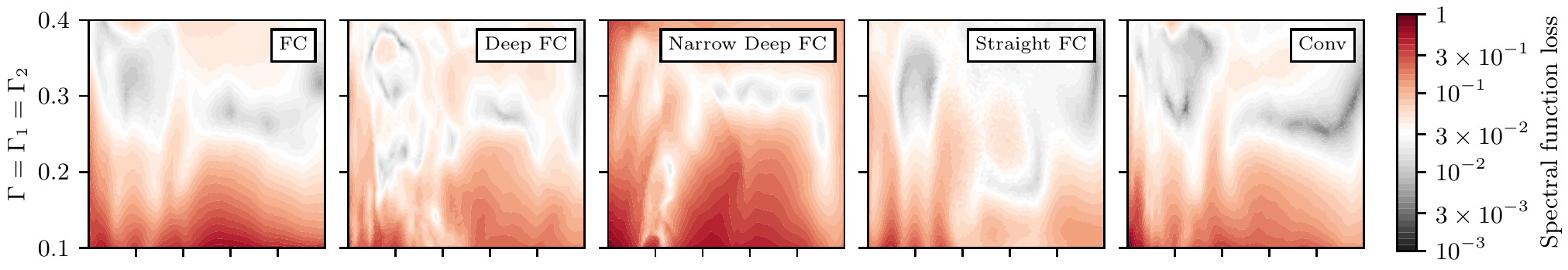}\\
  \includegraphics[width=1\textwidth]{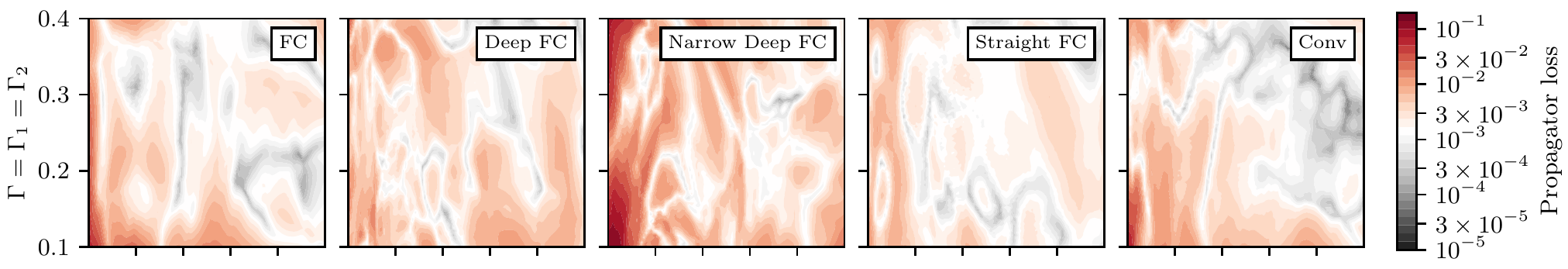}\\
  \includegraphics[width=1\textwidth]{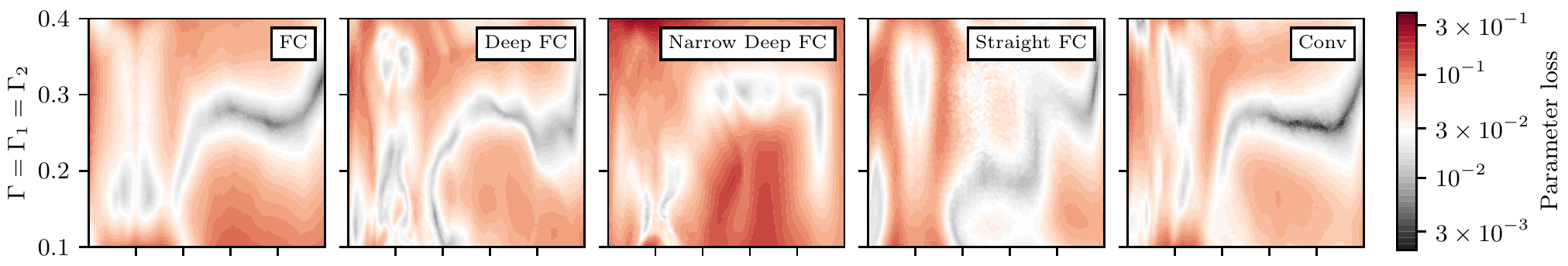}\\
  \includegraphics[width=1\textwidth]{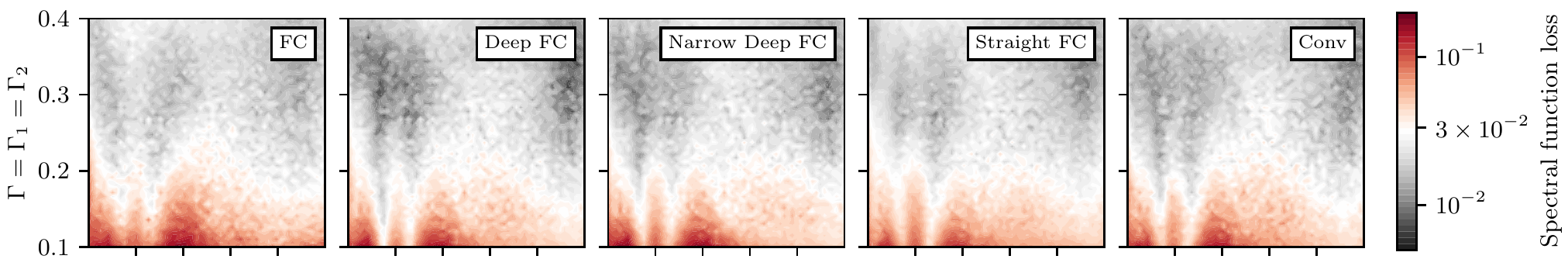}\\
  \includegraphics[width=1\textwidth]{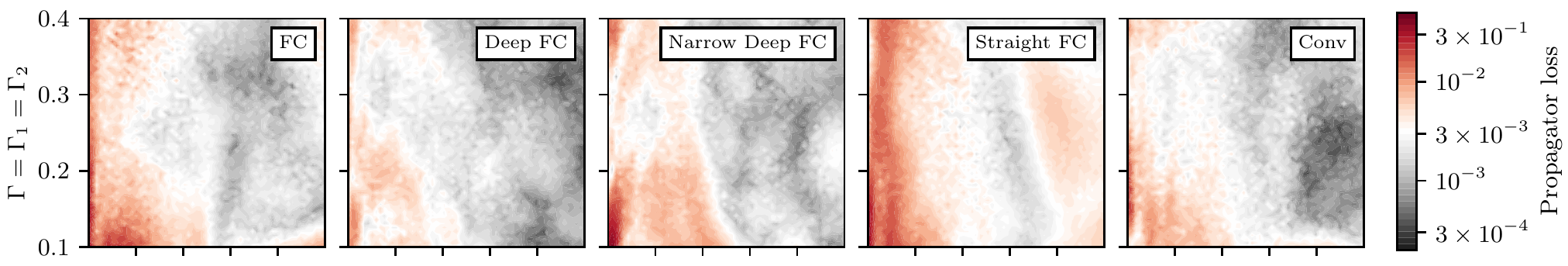}\\
  \includegraphics[width=1\textwidth]{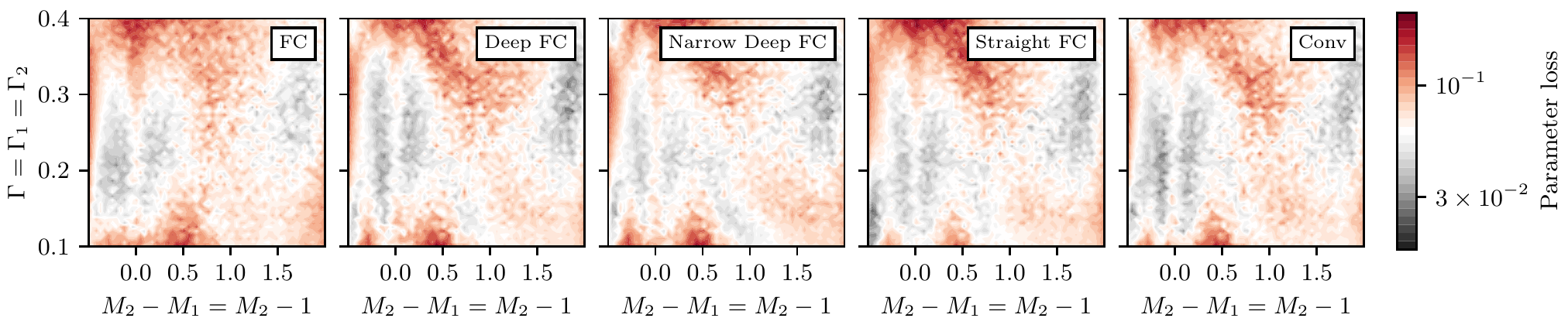} \caption{\textbf{Comparison
      of network architectures - } Contour plots of loss measures are
    shown for different net architectures. The upper three rows
    correspond to reconstructions of propagators with a noise width of
    $10^{-5}$, the lower ones with $10^{-3}$. The plots illustrate the
    loss measures in a hyperplane within the parameter space whose
    properties are described in \Cref{sec:mocktst}. The networks are
    trained with the parameter loss on the training set of volume Vol
    O. The contour plots show that the local minima are slightly
    different for small noise widths, whereas the global structures
    remain similar for all network architectures. These differences
    are caused by a slightly differing utilization of the limited
    number of hyperparameters. The differences between the network
    architectures become less visible for larger errors due to the
    growing severity of the inverse problem and a decreasing knowledge
    about the correct inverse transformations. Interestingly, the loss
    landscape of the convolutional neural network, which intrinsically
    operates on local structures, and of the fully connected networks
    are almost equal. The non-locality of the inverse integral
    transformation represents a possible reason for why the specific
    choice of the network structure is largely irrelevant. We conclude
    that the actual architecture is rather negligible in comparison to
    other attributes of the learning process, such as the selection of
    training data and the choice of the loss function.}
\label{fig:contournetarchs}
\end{figure*}

\begin{figure*}
  \includegraphics[width=1\textwidth]{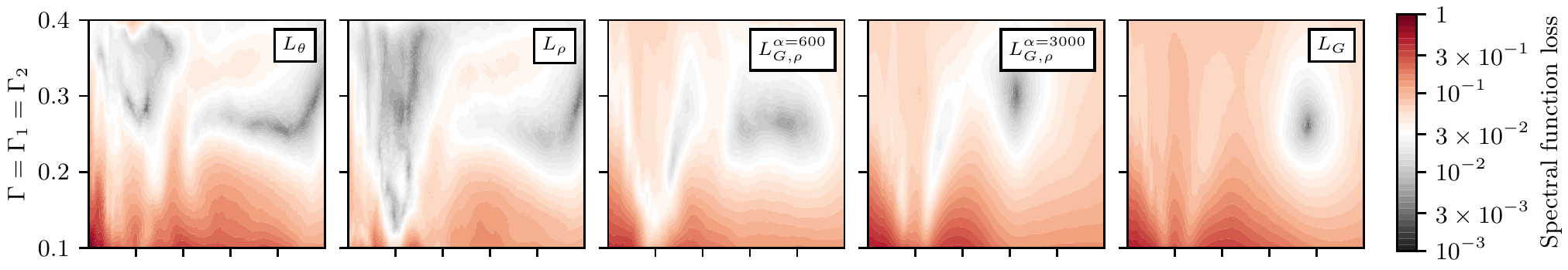}\\
  \includegraphics[width=1\textwidth]{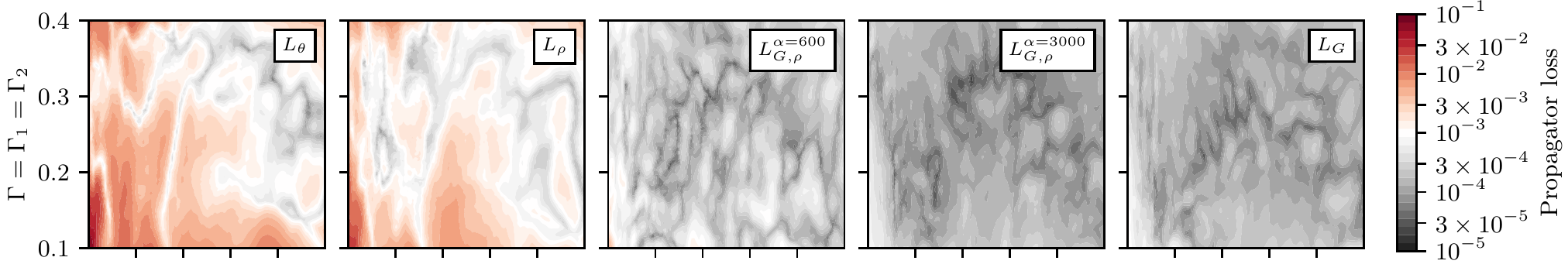}\\
  \includegraphics[width=1\textwidth]{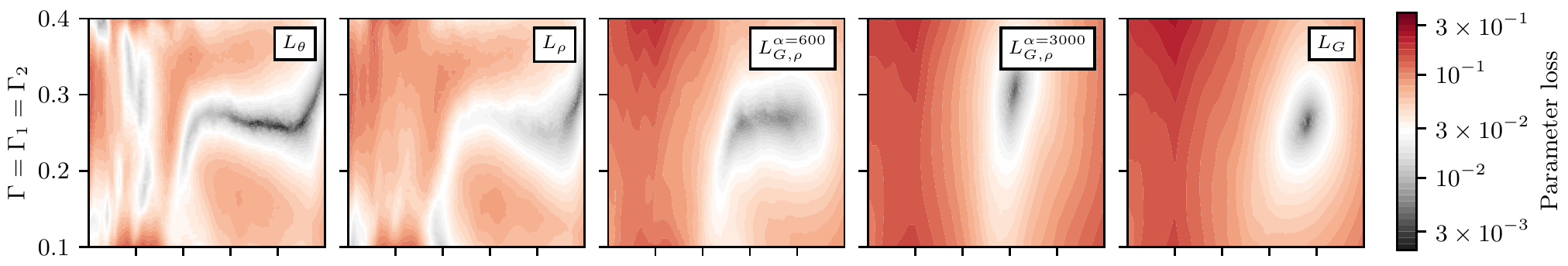}\\
  \includegraphics[width=1\textwidth]{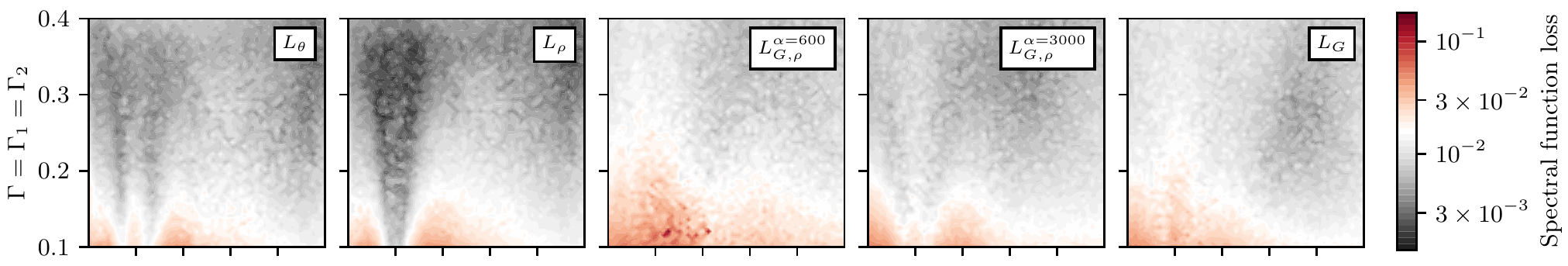}\\
  \includegraphics[width=1\textwidth]{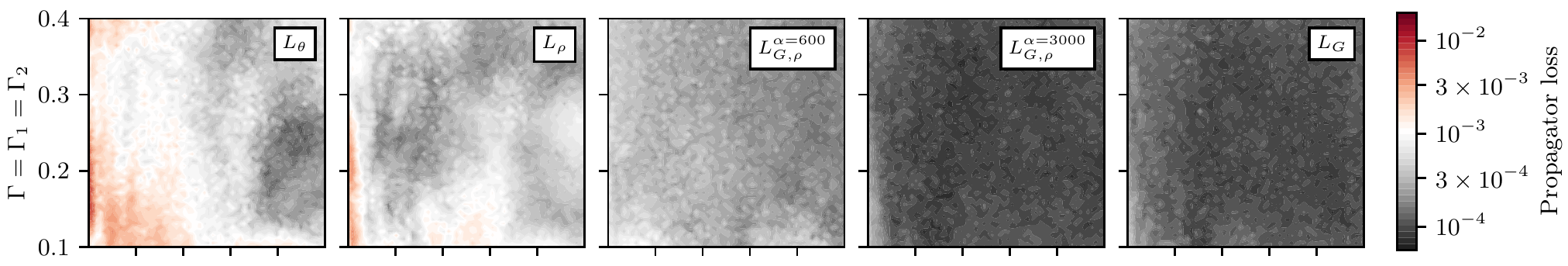}\\
  \includegraphics[width=1\textwidth]{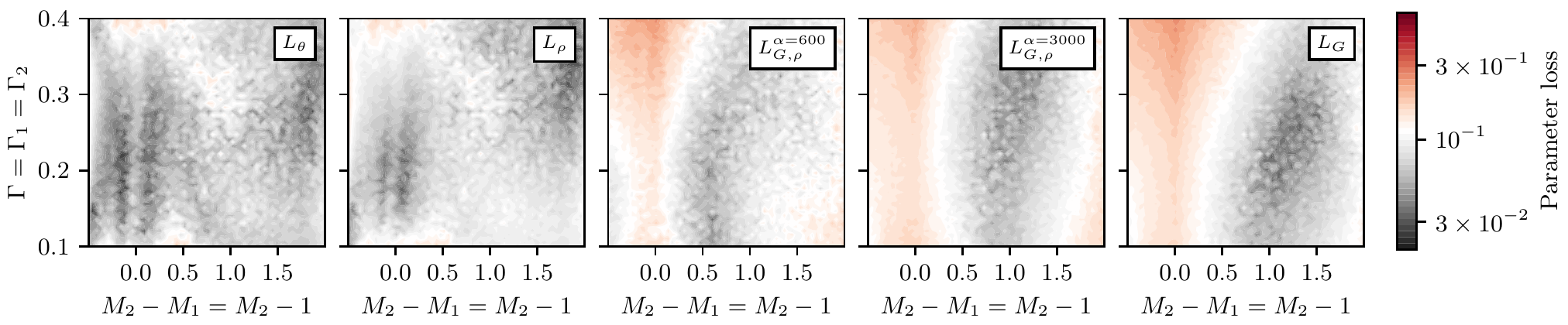} \caption{\textbf{Comparison
      of loss functions -} Contour plots of loss measures are
    illustrated in the same manner as in \Cref{fig:contournetarchs},
    but with a comparison o different loss functions. The considered
    loss functions are introduced in \Cref{sec:trainingstrategy}. The
    results are based on the Conv PaNet that is trained on volume Vol
    O. The optima in the loss function differ and, consequently, lead
    to different mean squared errors for the different measures. It is
    interesting that the network with the pure propagator loss
    function leads to a rather homogeneous propagator loss
    distribution. In contrast, the networks with the pure parameter
    and the pure spectral function loss do not result in homogeneous
    distributions for their corresponding loss function. The large set
    of nearly equal propagators for different parameters explains this
    observation. It confirms also once more the necessity of
    approaches that can be trained using loss functions with access to
    more information than just the reconstructed
    propagator.}\label{fig:contourloss}
\end{figure*}

\begin{figure*}
  \includegraphics[width=1\textwidth]{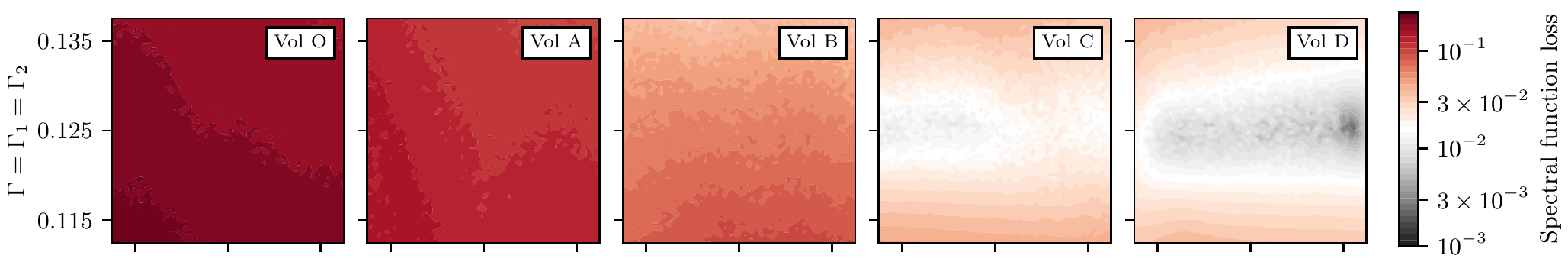}\\
  \includegraphics[width=1\textwidth]{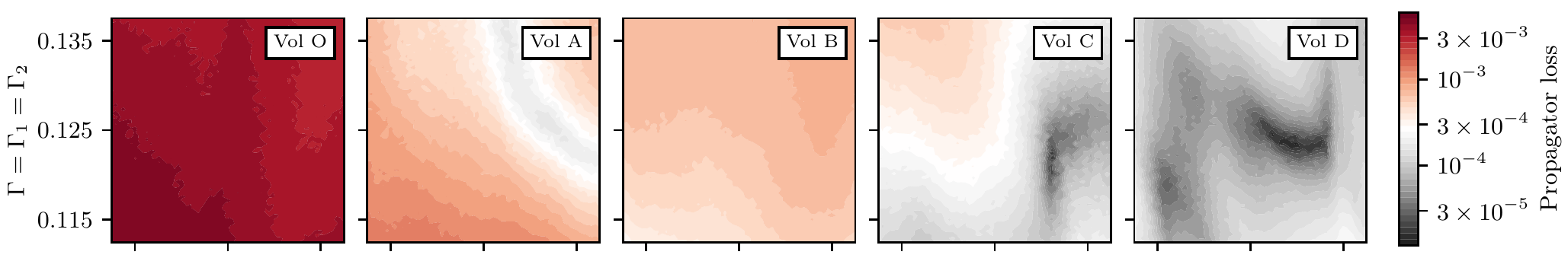}\\
  \includegraphics[width=1\textwidth]{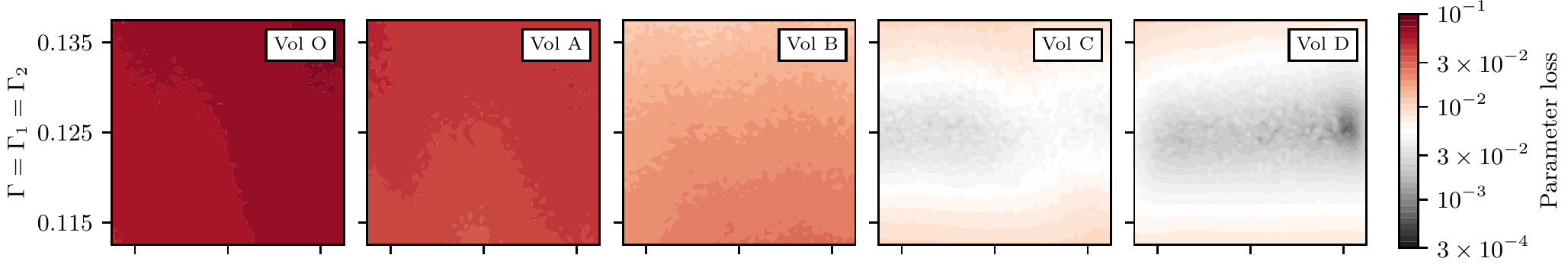}\\
  \includegraphics[width=1\textwidth]{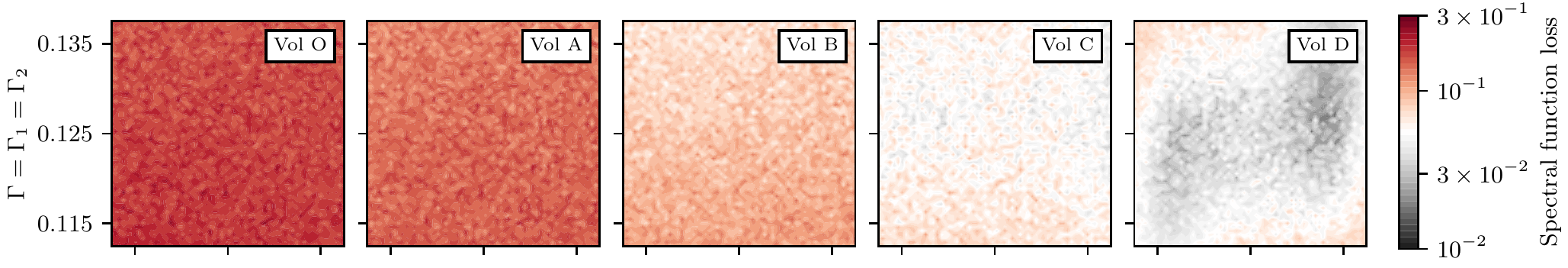}\\
  \includegraphics[width=1\textwidth]{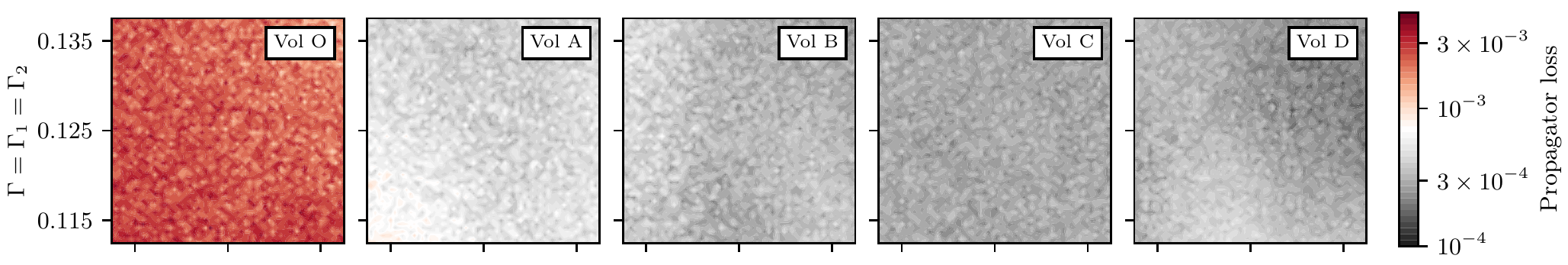}\\
  \includegraphics[width=1\textwidth]{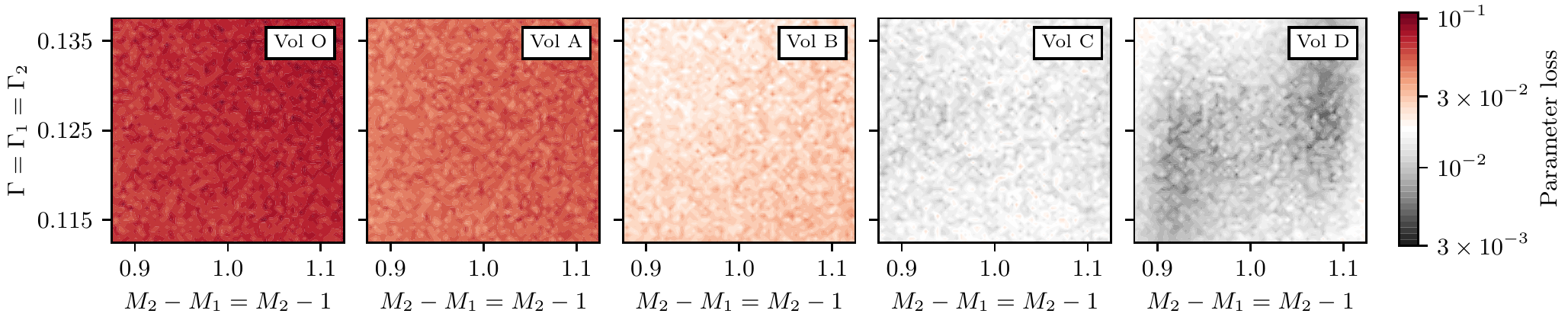}
  \caption{\textbf{Analysis of prior information (parameter space of
      the training data) and of local differences in the severity of
      the inverse problem -} The evolution of the landscape of
    different loss measures is shown for Conv PaNets that are trained
    on different parameter spaces. All contour plots are based on the
    same section of the parameter space, namely the range that is
    spanned by volume D. The upper three and lower rows correspond
    again to reconstructions of propagators with noise widths
    $10^{-5}$ and $10^{-3}$. The gradual reduction of the parameter
    space allows the analysis of different levels of complexity of the
    problem. A general improvement of performance can be observed
    besides a shift of the global optima. The more homogeneous loss
    landscape demonstrates that the problem of a different severity of
    the inverse problem is still present, but damped.}
	\label{fig:contourvol}
\end{figure*}

\begin{figure*}
	\includegraphics[width=0.48\textwidth]{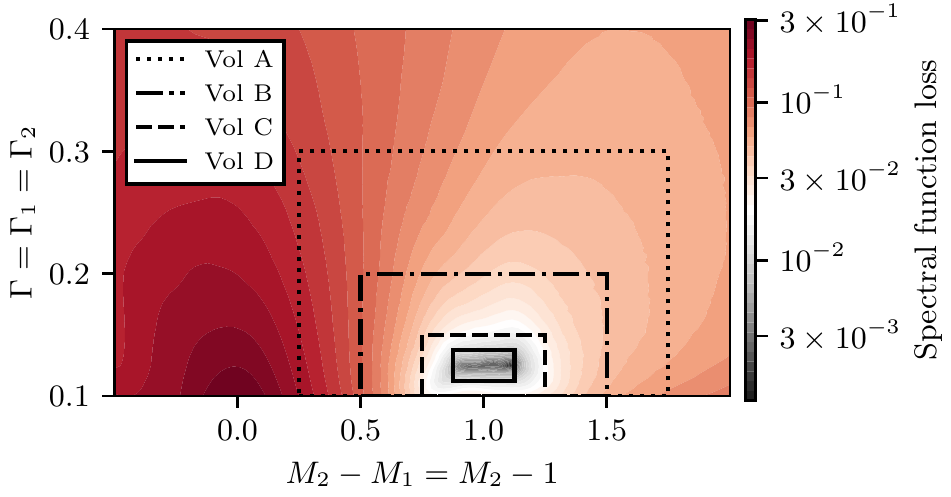} \includegraphics[width=0.48\textwidth]{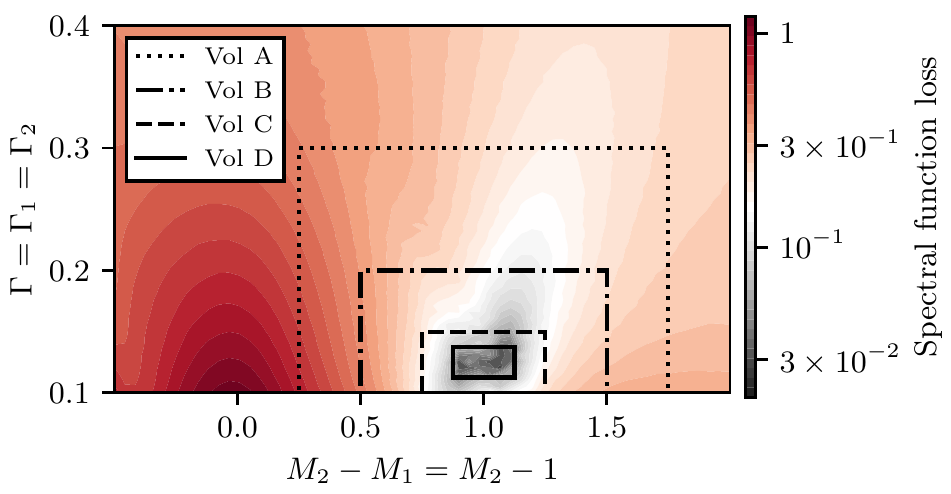}
	\includegraphics[width=0.48\textwidth]{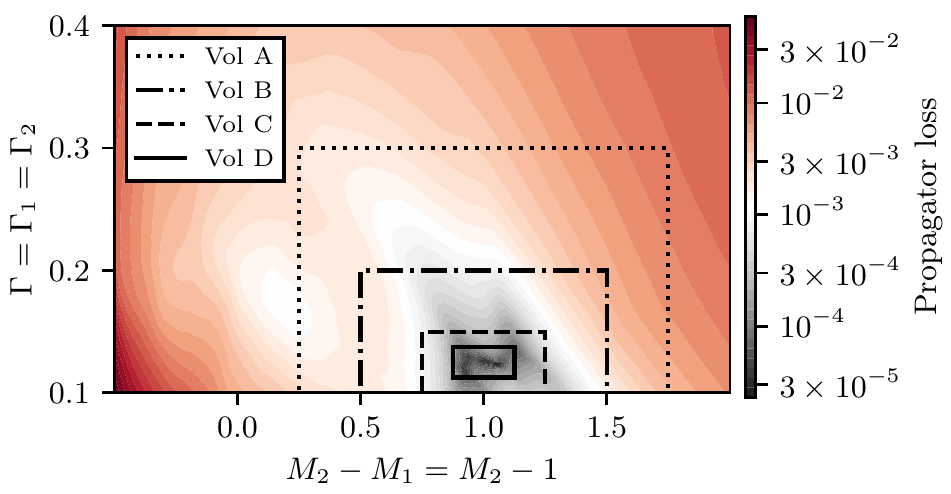}
	\includegraphics[width=0.48\textwidth]{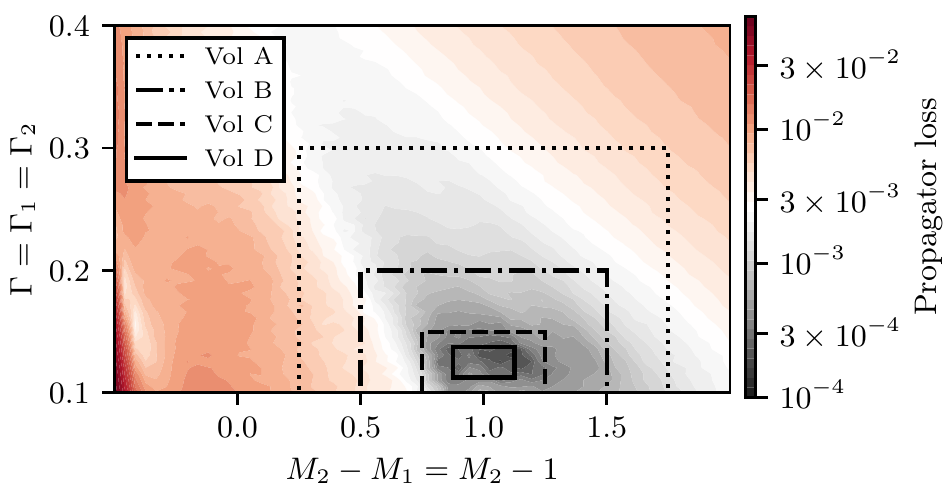}
	\includegraphics[width=0.48\textwidth]{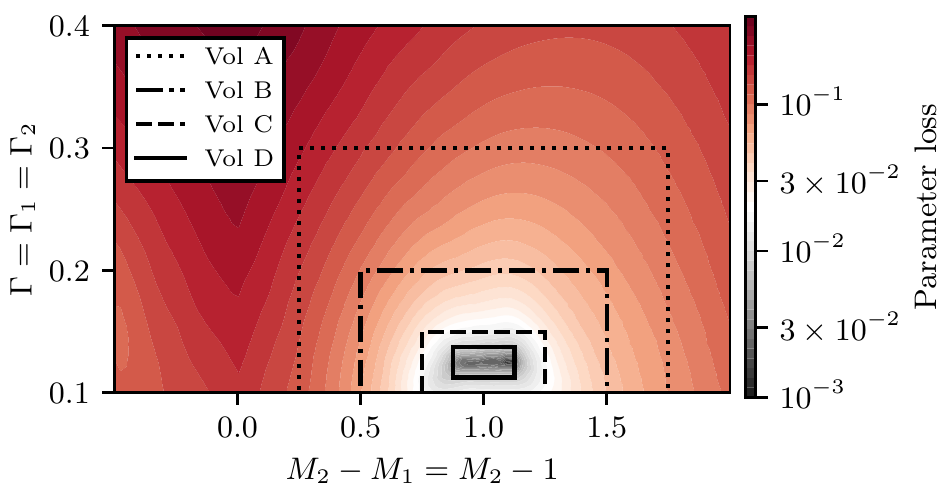}
	\includegraphics[width=0.48\textwidth]{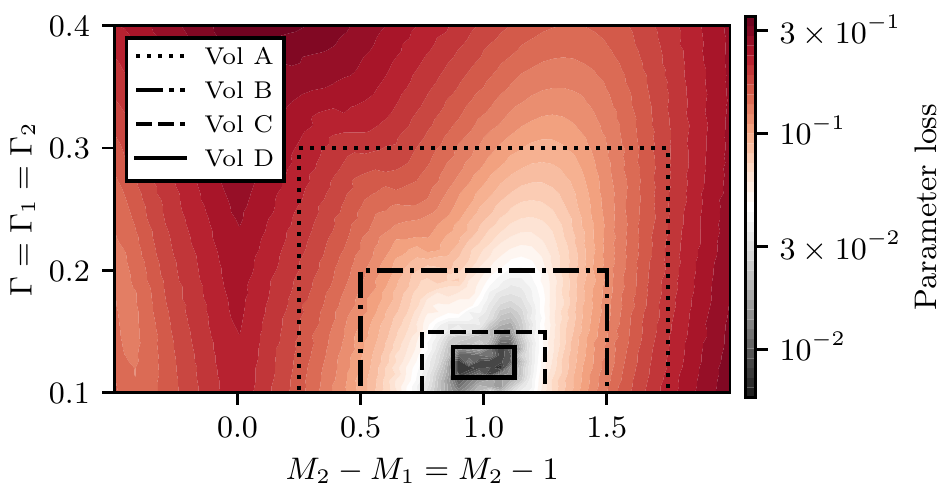}
	\caption{\textbf{Performance outside of the training region -}
          Performance of the Conv PaNet trained on the smallest volume
          Vol D for data that lies outside of the training region with
          a noise width of $10^{-5}$ (left column) and a noise width
          of $10^{-3}$ (right column). As expected, the prediction
          quality decreases with distance from the boundaries of Vol
          D. However, we emphasise that there is no immediate sharp
          transition at the boundary. Instead, we observe at first
          only a gradual decrease of the prediction quality,
          indicating that the network can generalise slightly beyond
          the trained region to varying degrees, depending on which
          parameters and error metrics are considered.}
	\label{fig:outofparameterspace}
\end{figure*}

\begin{figure*}
  \includegraphics[width=1\textwidth]{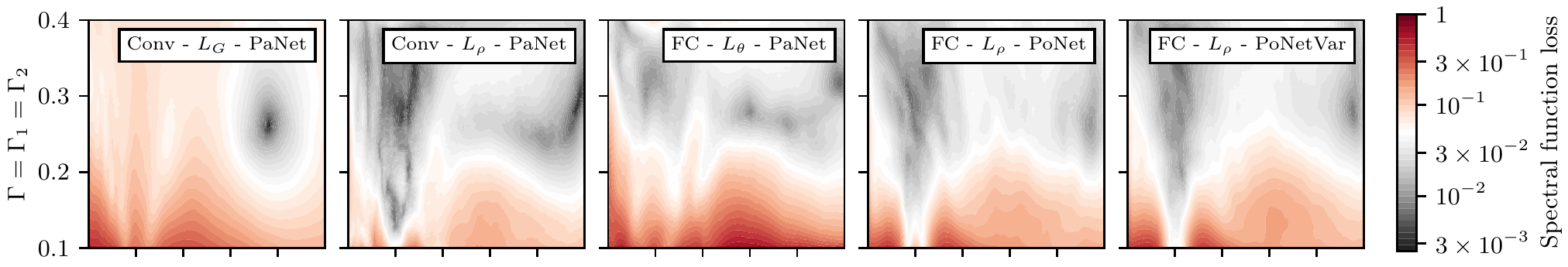}\\
  \includegraphics[width=1\textwidth]{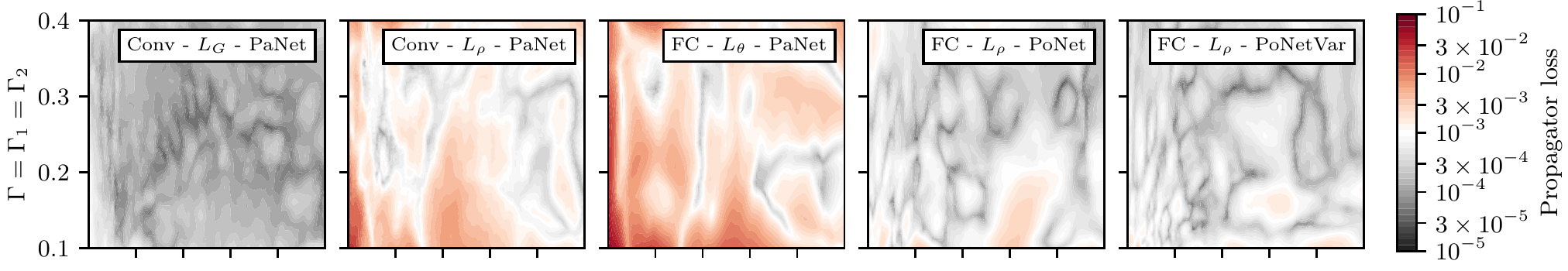}\\
  \includegraphics[width=1\textwidth]{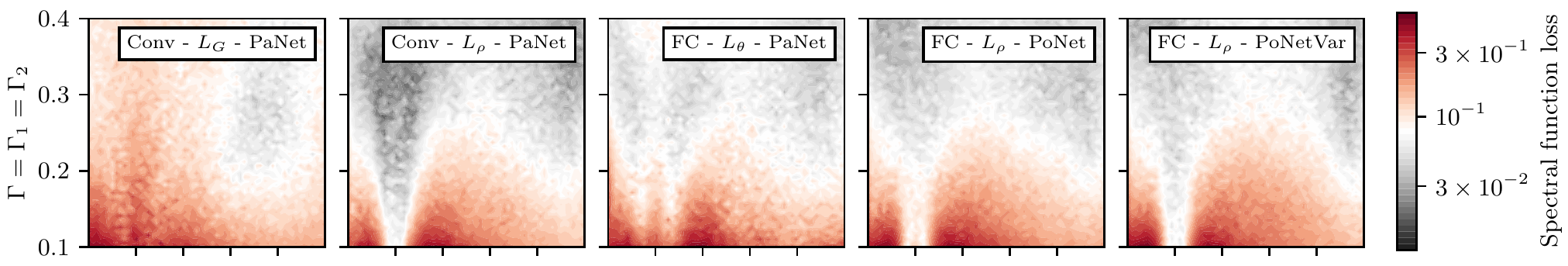}\\
  \includegraphics[width=1\textwidth]{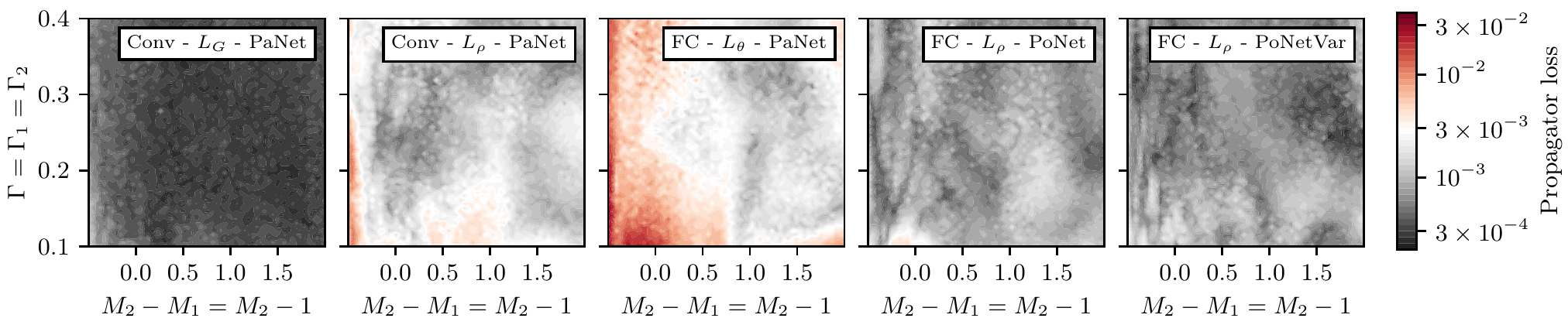}
  \caption{\textbf{Comparison of the parameter net and the point net
      -} Root-mean-squared-deviations are compared between the
    parameter net and the point net, trained on two Breit-Wigner like
    structures (PoNet) and trained on a variable number of
    Breit-Wigners (PoNetVar), with respect to different loss
    functions. The two upper rows correspond to results from input
    propagators with a noise width of $10^{-5}$ and the two lower ones
    with a noise width of $10^{-3}$. Problems concerning a varying
    severity of the inverse problem and concerning an information loss
    caused by the additive noise remain independent of the chosen
    basis for the representation of the spectral function.}
	\label{fig:contourpoint}
\end{figure*}

\begin{table*}
	\begin{tabular}{p{1.3cm}p{0.1cm}p{8cm}p{0.1cm}p{1.5cm}}
		Name & & CenterModule & & Number of parameters\\\hline\hline
		FC & &  FC(6700) $\Rightarrow$ ReLU $\Rightarrow$ FC(12168) $\Rightarrow$ ReLU $\Rightarrow$ FC(1024) & & $95\times10^6$\\\hline
		Deep FC & &  FC(512) $\Rightarrow$ ReLU $\Rightarrow$ FC(1024) $\Rightarrow$ ReLU $\Rightarrow$ (FC(4056) $\Rightarrow$ ReLU$)^3$ $\Rightarrow$ (FC(2056) $\Rightarrow$ ReLU$)^2$ & & $50\times10^6$\\\hline
		Narrow Deep FC & & FC(512) $\Rightarrow$ ReLU $\Rightarrow$ (FC(1024) $\Rightarrow$ ReLU$)^3$ $\Rightarrow$ (FC(2056) $\Rightarrow$ ReLU$)^5$ $\Rightarrow$ (FC(1024) $\Rightarrow$ ReLU$)^3$ $\Rightarrow$ FC(512) $\Rightarrow$ ReLU $\Rightarrow$ FC(256)  & & $96\times10^6$\\\hline
		Straight FC & &  (FC(4112) $\Rightarrow$ BatchNorm1D $\Rightarrow$ ReLU $\Rightarrow$ Dropout(0.2)$)^7$ & & $102\times10^6$\\\hline
		Conv & &  Conv(64, 10) $\Rightarrow$ ReLU $\Rightarrow$ Conv(256, 10) $\Rightarrow$ ReLU $\Rightarrow$ (FC(4096) $\Rightarrow$  ReLU$)^2$ $\Rightarrow$ FC(1024) & & $41\times10^6$\\
	\end{tabular}
	\caption{Details on the implemented network architectures. The
          argument of FC denotes the number of output neurons. The
          numbers in the argument of Conv correspond to the number of
          output channels and to the kernel size. The general setup
          is: Input(100) $\Rightarrow$ ReLU $\Rightarrow$ CenterModule
          $\Rightarrow$ ReLU $\Rightarrow$ FC(3/6/9/500) $\Rightarrow$
          Output, where the CenterModule is given along with the
          associated name in the Table. The size of the output layer
          is determined by the use of a parameter net or a point net
          and the considered number of Breit-Wigners.}
	\label{tab:netarchitecures}
\end{table*}

\begin{table*}
	\begin{tabular}{p{1.7cm}p{1.4cm}p{3.2cm}p{2.8cm}p{1.9cm}p{2.7cm}p{1.7cm}p{1.2cm}}
		Figure & Network type & Architecture & Loss function & Training set & Test set & Noise width & Number of BWs \\\hline\hline
		\Cref{fig:numbwimpact} & PaNet &  FC & $L_{\theta}$ & Vol O & Noise samples on same propagator & $10^{-3}$ & 1-3 \\
		\Cref{fig:histonetarchs} & PaNet &  Various (a) / Conv (b) & $L_{\theta}$ (a) / Various (b) & Vol O & Vol O & $10^{-3}$ / $10^{-5}$ & 2 \\
		\Cref{fig:volumeimpact} & PaNet &  Conv & $L_{\theta}$ & Various & Noise samples on same propagator & $10^{-3}$ & 2 \\
		\Cref{fig:comparisonvolume} & PaNet &  Conv / Conv PP & $L_{\theta}$ & Various & Vol D & Various & 2 \\
		\Cref{fig:comparisonoutofrange} & PaNet &  Conv & $L_{\theta}$ & Vol D & Various & Various & 2 \\
		\Cref{fig:comppoint} & Various &  FC & Various & Vol O & Vol O & Various & 1-3 \\
		\Cref{fig:errorimpact} & PaNet &  Conv & $L_{\theta}$ & Vol B & Noise samples on same propagator & Various & 2 \\
		\Cref{fig:comparisonall} & PaNet &  FC / FC PP & $L_{\theta}$ & Vol O & Vol O & Various & 1-3 \\
		\Cref{fig:comparisoncontourall} & PaNet &  FC & $L_{\theta}$ & Vol O & Contour - Vol O & $10^{-3}$ & 2 \\
		\Cref{fig:brcomparison} & PaNet &  Conv & $L_{\theta}$ & See~\Cref{tab:volumesbrcompairson} & Specific sets & $10^{-3}$ & 1-3 \\
		\Cref{fig:contournetarchs} & PaNet &  Various & $L_{\theta}$ & Vol O & Contour -  Vol O & $10^{-3}$ / $10^{-5}$ & 2 \\
		\Cref{fig:contourloss} & PaNet &  Conv & Various & Vol O & Contour - Vol O & $10^{-3}$ / $10^{-5}$ & 2 \\
		\Cref{fig:contourvol} & PaNet &  Conv & $L_{\theta}$ & Various & Contour - Vol D & $10^{-3}$ / $10^{-5}$ & 2 \\
		\Cref{fig:outofparameterspace} & PaNet &  Conv & $L_{\theta}$ & Vol D & Contour - Vol O & $10^{-3}$ / $10^{-5}$ & 2 \\
		\Cref{fig:contourpoint} & Various &  Conv / FC & Various & Vol O & Contour - Vol O & $10^{-3}$ / $10^{-5}$ & 2	
	\end{tabular}
	\caption{List of figures that contains details about the
          training of the different networks and about the dataset
          used for evaluation/validation.}
	\label{tab:networktrainingdetails}
\end{table*}

\bibliographystyle{apsrev4-1} \bibliography{BibCollection}

%%%%%%%%%%%%%%%%%%%%%%%%%%%%%%%%

\end{document}